\documentclass[prb,reprint,superscriptaddress,amsmath,amssymb]{revtex4-1}
\usepackage{graphicx}% Include figure files
\usepackage{dcolumn}% Align table columns on decimal point
\usepackage{bm}% bold math
\usepackage{color}
\usepackage{url}

\begin{document}

\title{$^{59}$Co NMR evidence for charge and orbital order in the kagom\'{e}
like structure of Na$_{2/3}$CoO$_{2}$}

\author{I.R.~Mukhamedshin}
\email{Irek.Mukhamedshin@ksu.ru}
\affiliation{Laboratoire de Physique des Solides, UMR 8502, Universit\'e Paris-Sud, 91405 Orsay, France, EU}
\affiliation{Institute of Physics, Kazan University, 420008 Kazan, Russia}
\author{H.~Alloul}
\affiliation{Laboratoire de Physique des Solides, UMR 8502, Universit\'e Paris-Sud, 91405 Orsay, France, EU}

\begin{abstract}
We report a complete set of $^{59}$Co NMR data taken on the $x=2/3$ phase of
sodium cobaltates Na$_{x}$CoO$_{2}$, for which we have formerly established
the in plane Na ordering and its three dimensional stacking from a
combination of symmetry arguments taken from Na and Co NQR/NMR data. Here we resolve all the parameters of the Zeeman and quadrupolar Hamiltonians for all cobalt sites in the unit cell and report the temperature dependencies of the NMR shift and spin lattice relaxation $T_{1}$ data for these sites. We confirm that three non-magnetic Co$^{3+}$ (Co1) are in axially
symmetric positions and that the doped holes are delocalized on the nine
complementary magnetic cobalt sites (Co2) of the atomic unit cell. The
moderately complicated atomic structure resumes then in a very simple
electronic structure in which the electrons delocalize on the Co2 kagom\'e
sublattice of the triangular lattice of Co sites. The observation of a single temperature dependence of the spin susceptibilities indicates that a single band picture applies, and that the magnetic properties are dominated by the static and dynamic electronic properties at the Co2 sites. We evidence that they display a strong in plane electronic anisotropy initially unexpected but which accords perfectly with an orbital ordering along the kagom\'e sublattice organization. These detailed data should now permit realistic calculations of the electronic properties of this compound in order to determine the incidence of electronic correlations.
\end{abstract}


% 76.60.-k Nuclear magnetic resonance and relaxation
% 71.27.+a Strongly correlated electron systems; heavy fermions
% 61.66.-f Structure of specific crystalline solids
% 71.28.+d Narrow-band systems; intermediate-valence solids
% 75.20.Hr Local moment in compounds and alloys; Kondo effect, valence fluctuations, heavy fermions
% 76.60.Gv Quadrupole resonance

\maketitle

\section{Introduction}

The cobaltates Na$_{x}$CoO$_{2}$ are layered oxyde materials somewhat
similar to the cuprates in as much as the charge doping of the CoO$_{2}$
layers is controlled on a large range by variation of the Na content. This
can be put in parallel with the doping of the cuprates by chemical
substitutions on the layers separating the CuO$_{2}$ planes. One significant
difference with the cuprates is that the Co of the CoO$_{2}$ plane are
ordered on a triangular lattice and not on a square lattice as for the CuO$%
_{2}$ plane of the cuprates. In this configuration the large crystal field
on the Co site favors a low spin state in which orbital degeneracy
influences significantly the electronic properties and may yield large
thermoelectric effects.\cite{TerasakiTEP} A rich variety of other physical
properties ranging from ordered magnetic states,\cite{Mendels05} high
Curie-Weiss magnetism and metal insulator transition,\cite{Foo}
superconductivity\cite{TakadaNature} \emph{etc} have then been observed on the
cobaltates.

In most cases the theoretical calculations considered to explain the
physical properties of cuprates and cobaltates have assumed uniform
delocalization of the carriers in the layers\cite{SinghPRB61} and the incidence of the coulomb potential of the ionic dopants has been generally considered as
unimportant.

Experimentally though, the disorder due to the dopants has been shown to
have an important incidence on the local electronic properties in the
cuprates and can blur the physical properties and may drive as well system
specific, that is non generic properties of the CuO$_{2}$ planes. For
instance static stripe charge organization of the CuO$_{2}$ planes have been
seen only on specific systems such as La$_{2-x}$Ba$_{x}$CuO$_{4}$ for which
static distortions lead to an orthorombicity of the structure.\cite{LaBaCuO}
Some anisotropic electronic properties of YBa$_{2}$Cu$_{3}$O$_{6+x}$
compounds might as well tentatively be associated with the incidence of the
ordered CuO chains which drive the doping of the CuO$_{2}$ planes.\cite{AndoPRL}

In the cobaltates, NMR experiments and structural investigations have given
evidence that for $x \geqslant 0.5$ a large interplay occurs between atomic
arrangements and electronic properties, as the Na are found to be ordered.%
\cite{Zandbergen,NaPaper,EPL2008,Roger,Shu,Huang09} It has been shown earlier that this ordering is associated quite systematically with cobalt charge disproportionation into $3^{+}$ and $\approx 3.5^{+}$ Co charge states.\cite{CoPaper,EPL2008,LangNFD,MHJulien075} As the cobalt ions in sodium cobaltates are in low spin configurations the Co$^{3+}$ has an electronic spin $S=0$ and appears to be inert magnetically, when compared to the other Co sites with higher charge state on which holes delocalize (formally Co$^{4+}$ should have $S=1/2$). However the actual Na atomic order, the organization of the Co sites and their electronic properties has not been so far determined experimentally, except for the case of $x=1/2$ where such a large charge differentiation does not occur.\cite{Bobroff05,Imai05} This did not therefore permit yet any theoretical calculations, even in the LDA approximation, based on an actual structure.

\begin{figure}[tbp]
\center
\includegraphics[width=0.8\linewidth]{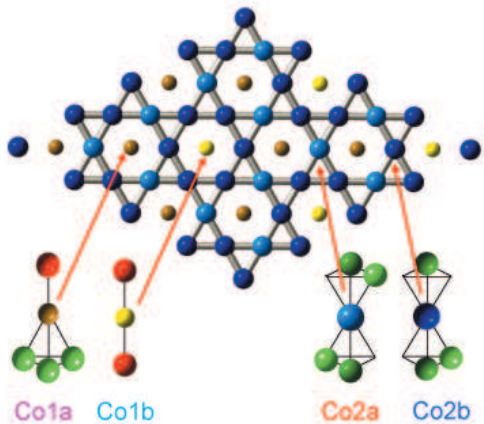}
\caption{(Color online) Two dimensional structure of the Co planes of Na$%
_{2/3}$CoO$_{2}$ compound and the Na environments of the four Co sites
differentiated in this structure are displayed. Cobalts in the Co1a and Co1b
positions have a threefold symmetry and one or two sodium ions in Na1
position (red balls) as nearest neighbors which drive them into a
non-magnetic Co$^{3+}$ charge state. In the Co2a and Co2b positions cobalts
have nearest neighbours sodium ions in Na2 positions (green balls). The
small difference of local environment of these two Co sites only yields
negligible distinction between their electronic properties. This results in
a 2D structure where these Co sites form a kagom\'{e} lattice.%
\protect\cite{EPL2009,H67NQRprb}}
\label{fig:StrucCoH67}
\end{figure}

We have however established for long that one specific structure of Na$_{x}$%
CoO$_{2}$ was quite stable for $x\approx 0.7$ and we could ascertain
recently that it corresponds to a $x=2/3$ atomic structure which could be
determined by combining NMR and NQR experiments.\cite{EPL2009} The Na
organization in this structure, which agrees with GGA calculations,\cite%
{Hinuma} consists of two Na on top of Co sites (the Na1 sites) and six on
top of Co triangles (the Na2 sites). This twelve Co unit cell and the
stacking between planes has been confirmed by x-ray diffraction experiments.%
\cite{H67NQRprb} It results in a differentiation of four cobalt sites in the
structure, two nearly non-magnetic Co$^{3+}$ and two more magnetic sites
constituting a kagom\'e sublattice of the triangular Co lattice on which the
holes are delocalized (see Fig.~\ref{fig:StrucCoH67}).

Let us point out that a conflicting suggestion concerning the structure of this phase has been proposed in Ref.~\onlinecite{TaiwanPRB2010}. In Appendix~\ref{TaiwaneseAreWrong} we demonstrate that our data are quite incompatible with this alternative structure, and that nothing allows so far  to establish the occurrence of oxygen vacancies in our samples.

In the present paper we shall present a further important experimental step
which consists in the determination of a full set of $^{59}$Co NMR data
which, together with NQR data, allow us to unravel the electronic
properties of the four considered Co sites. We confirm that
the electronic structure resumes in two type of Co sites, the non-magnetic
Co1 sites with charge state 3$^{+}$ and the Co2 sites on which holes
delocalize. The magnetic properties sensed by all four sites have a single $T$ dependent behaviour which confirm a single band picture suggested initially.\cite{CoPaper} We evidence unexpected large anisotropies of the magnetic NMR shift and spin lattice relaxation of the Co2 sites, which have direct implications on the electronic structure of the CoO$_{2}$ plane. These
results should trigger realistic calculations to explain the correlated
electronic structure of these compounds and their evolution with hole doping.

The paper is organized as follows. Sections ~\ref{Samples}-\ref{SectionCoT1} contain all experimental aspects of the paper, while in Sec.~\ref{SectionDiscussion} we collected and discussed all the results which are relevant for further theoretical considerations.

In Sec.~\ref{Samples} we recall briefly the methods used to synthesize single phase samples for $x=2/3$ and to orient the samples in an applied field. More details on the NMR tests of the quality of the sample alignment are given in Appendix~\ref{AppendixPowdAlign}. In Sec.~\ref{SectionCoSpect} we recall some basics of the NMR spectroscopy of nuclear spins $I>1/2$ in presence of electric field gradients (EFG) which induce quadrupole effects. We report then the $^{59}$Co NMR spectra detected in Na$_{x}$CoO$_{2}$ samples when the field is
applied in both directions $H\parallel c$ and $H\perp c$, and the
experimental procedures used to separate the contributions of the different
sites to the NMR spectrum. We describe then the simulations of the NMR
spectra which allowed us to evidence the large in plane anisotropy of the
NMR shift of the magnetic sites.

In Sec.~\ref{TdepCoshift} we present the $T$ dependencies of the NMR shifts
of all Co sites in the structure and compare them to that observed on the Na
NMR (technical details are reported in Appendix~\ref{AppendixCo2XY}). This
allowed us to separate the $T$ dependent spin contribution to the NMR shifts from a $T$ independent term that we assign to an orbital contribution. We evidence that both terms are non-axial on the Co2 sites.

In Sec.~\ref{SectionCoT1} the technical aspects of the measurements of the $%
^{59}$Co NMR spin lattice relaxation are presented and the data obtained on
the two types of cobalt sites are analyzed by comparing the Co $T_{1}$ data
to that taken formerly on $^{23}$Na. This permits us as well to separate the
spin and orbital contributions to $(T_{1}T)^{-1}$. The anisotropies of the
data for the spin contributions can be explained by those obtained for the
hyperfine coupling. These results allow us as well to demonstrate that the spin and orbital contributions to $(T_{1}T)^{-1}$ are at least an order of magnitude larger for Co2 than for Co1 sites.

In Sec.~\ref{SectionDiscussion} we then discuss the ensemble of results on
the EFG, the NMR shifts and $(T_{1}T)^{-1}$ data. While so far the
qualitative aspects of the data had lead us to consider that the Co1 sites
were Co$^{3+}$, the measured parameters allow us to get a little bit further
and to obtain an upper limit on the charge occupancy of the Co1
sites. More importantly this allows us to consider the origin of the
anisotropy on the Co2 sites and to conclude that it results from the
distribution of holes between the axial $a_{1g}$ orbitals and the in plane $%
e_{g}^{\prime}$ orbitals on the Co2 site.

\section{Sample preparation and experimental techniques}

\label{Samples}

Experiments on sodium cobaltates Na$_{x}$CoO$_{2}$ have been performed on
both powders or single crystals. Usually, NMR experiments, which provide information on the bulk of the samples are expected to bring more information on single crystals than in powder samples. However this has not be proven to be the case so far.\cite{ImaiPRL1, MHJulien075} Indeed, we have shown that
homogeneous single phase powder samples can be synthesized and can be well
controlled by x-rays and NMR.\cite{EPL2008} Also the penetration of rf
magnetic field into conducting cobaltates single crystals is limited by the
skin effect. Thus for our study we used powder samples. However, as
they consist of particles oriented randomly, the resulting angular
distribution of quadrupole splittings introduces as well some difficulties
in the analysis of the NMR spectra.

\subsection{Samples preparation}

The methods which have been used to synthesize reproducibly single-phase
powder samples with sodium content x=0.67 have been reported in Ref.~%
\onlinecite{H67NQRprb}. As described there, we used three different routes
to synthesize homogeneous single phase samples of Na$_{2/3}$CoO$_{2}$
compounds:

(1) direct synthesis from a stoichiometric composition of Co$_{3}$O$_{4}$
and Na$_{2}$CO$_{3}$.

(2) from a mixture of cobaltates with calibrated compositions synthesized
previously (such as Na$_{1/2}$CoO$_{2}$ and Na$_{0.71}$CoO$_{2}$ \cite%
{EPL2008}) taken in a proper ratio.

(3) by de-intercalation of Na from Na$_{0.71}$CoO$_{2}$ by annealing it at
700${^{\circ }}$C - out of its own stability temperature range.

Whatever the synthesis procedure used, the X-ray spectra of these Na$_{2/3}$%
CoO$_{2}$ samples displayed the same $c$ axis parameter and diffraction
spectra, including the satellite Bragg peaks due to Na order.\cite{H67NQRprb}
However we noticed an important difference between these materials regarding
the possibility to orient them in a magnetic field.

\subsection{Powder sample alignment}

\label{Powder sample alignment}

As the room $T$ susceptibility of sodium cobaltates is known to be
anisotropic,\cite{Wang} we have used this anisotropy to align the single
crystallites of powder samples in a 7~T field, by mixing the samples with
Stycast 1266 epoxy resin which cured in the field. Thus in our samples
crystallites $c$ axes were aligned in the same direction, but $ab$ planes of
different crystallites are randomly distributed.\cite{Egorov} To be
successful, this procedure requires a powder in which individual grains
should be single crystallites and should not form clusters. Also the shape
and packing of the grains should not prevent them rotating freely in the
magnetic field.

The best aligned samples were obtained from powders which were directly
synthesized along route (1) with grain sizes of 50-100 microns, but a residual unreacted Co$_{3}$O$_{4}$ could not be avoided in many cases and the corresponding $^{59}$Co NMR signal of Co$_{3}$O$_{4}$ could be seen in the spectra. The powders obtained along route (3) also usually align well in the magnetic field. Those synthesized along route (2) were most difficult and sometimes impossible to align. In such a synthesis the small grains are probably randomly welded to each other, so that the final powder does not consist of single crystallites.

For a perfect alignment the samples should look like a single crystal in the
$c$ direction but with full $ab$ plane disorder. Details on the NMR tests
of the quality of the sample alignment are given in Appendix~\ref%
{AppendixPowdAlign}.

Here we should mention that the Stycast perfectly protects the powder from
environment (water) influence.\cite{4phasesNQR} For this Na concentration
the samples prepared 6 years ago did not show any changes in the NMR spectra.

\section{$^{59}$C\lowercase{o} NMR spectra in the N\lowercase{a}$_{2/3}$C\lowercase{o}O$_{2}$}

\label{SectionCoSpect}

\subsection{NMR Background}

Generally in solid state NMR an atomic nucleus with spin $\vec{I}$ and
quadrupole moment $Q$ has its spin energy levels determined by the Zeeman
interaction with a external magnetic field $\vec{H}_{0}$ and the quadrupolar
interaction with the electric field gradient (EFG) on the nucleus site.
Therefore the Hamiltonian consists of two parts - Zeeman Hamiltonian $%
\mathcal{H}_{Z}$ and quadrupolar Hamiltonian $\mathcal{H}_{Q}$, and can be
written\cite{Abragam,Slichter}:
\begin{equation}
\mathcal{H}=\mathcal{H}_{Z}+\mathcal{H}_{Q}=-\gamma \hbar \vec{I}(1+\hat{K})%
\vec{H_{0}}+\frac{eQ}{2I(I-1)}\vec{I}\hat{V}\vec{I},  \label{eq:Hamiltonian}
\end{equation}%
where $\gamma $ is gyromagnetic ratio.

The physical properties of the studied compound are hidden in the two
tensors in this Hamiltonian - magnetic shift tensor $\hat{K}$ and the EFG
tensor $\hat{V}$. The principal axes of both tensors are associated with the
local structure and assuming that the principal axes (\emph{X,Y,Z}) of both
tensors coincide, the Zeeman and quadrupolar Hamiltonians can be re-written:
\begin{eqnarray}
&&\mathcal{H}_{Z} = -\gamma \hbar \sum_{\alpha }I_{\alpha }(1+K_{\alpha
\alpha })H_{0\alpha }; \nonumber \\
&&\mathcal{H}_{Q} = \frac{eQ}{2I(I-1)}\sum_{\alpha }V_{\alpha \alpha
}I_{\alpha }^{2} \label{eq:Hamilt2}\\
&&\;\;\;\;=\frac{h\nu _{Z}}{6}\left[ 3I_{Z}^{2}-I(I-1)\right] +(\nu _{X}-\nu
_{Y})(I_{X}^{2}-I_{Y}^{2})),\nonumber
\end{eqnarray}%
where $\alpha =X,Y,Z$, and the three quadrupolar frequencies
\begin{eqnarray*}
\nu_{\alpha}=\frac{3eQV_{\alpha \alpha }}{\left[2I(2I-1)h\right]}
\end{eqnarray*}%
are linked by Laplace equation $\sum_{\alpha }V_{\alpha \alpha
}=\sum_{\alpha }\nu _{\alpha }=0$. Therefore it is more common to use two
parameters - the quadrupolar frequency $\nu _{Q}=\nu _{Z}$ corresponding to
the largest principal axis component $V_{ZZ}$ of the EFG tensor and the
asymmetry parameter $\eta =(V_{XX}-V_{YY})/V_{ZZ}$ (here the principal axes
of the EFG tensor are chosen following $\left\vert V_{ZZ}\right\vert \geq
\left\vert V_{YY}\right\vert \geq \left\vert V_{XX}\right\vert $).

The nuclear spin of $^{59}$Co is $I=7/2$ and therefore for a given direction
of applied magnetic field $H_{0}$ relative to the crystallite the NMR
spectrum for a single Co site consists of 7 lines - a central line which
corresponds to the $-\frac{1}{2}\leftrightarrow \frac{1}{2}$ transition and
6 satellites corresponding to the other $m\leftrightarrow (m-1)$ transitions. The position of the central line is determined by the applied field and the
values of the magnetic shift $K$ and second order quadrupolar perturbation.%
\cite{Abragam} The distance between the two satellite lines $%
m\leftrightarrow (m-1)$ and $-m\leftrightarrow -(m-1)$ depends on the
orientation of the external field with respect to the principal axis of EFG
tensor described by the spherical angular coordinates $\theta$ and $%
\varphi$ and can be expressed as:\cite{Abragam}
\begin{equation}
\Delta \nu =\nu _{Q}(m-\frac{1}{2})(3\cos ^{2}\theta -1+\eta \sin ^{2}\theta
\cos 2\varphi).  \label{eq:SatelDist}
\end{equation}

\subsection{Experimental techniques}

The NMR measurements were done using a home-built coherent pulsed NMR
spectrometer. NMR spectra were taken \textquotedblleft point by point
\textquotedblright with a $\pi /2-\tau -\pi /2$ radio frequency (rf) pulse
sequence by varying the magnetic field in equal steps. The minimum practical
$\tau $ values used in our experiments was 7~$\mu s$. The usual $\pi /2$
pulse length was 2~$\mu s$.

Experimentally the value of the magnetic shift (in \%) for a given point in
the spectrum either in frequency domain or in the field domain can be
calculated using:
\begin{equation}
K=\frac{\nu -\nu _{ref}}{\nu _{ref}}=\frac{B_{ref}-B}{B},  \label{eq:Kdef}
\end{equation}%
where $\nu _{ref}$ and $B_{ref}$ are reference frequency and field values,
respectively, which are connected by the relation $\nu _{ref}=(\gamma /2\pi
)B_{ref}$. In this work $^{23}$Na shifts are given with respect to the $%
^{23} $Na resonance in a NaCl water solution. The reference for $^{59}$Co
shift is given for the $\gamma /2\pi =$ 10.054~MHz/T value.

\subsection{Decomposition of the spectra in slow and fast relaxing components}

\begin{figure}[tbp]
\center
\includegraphics[width=1\linewidth]{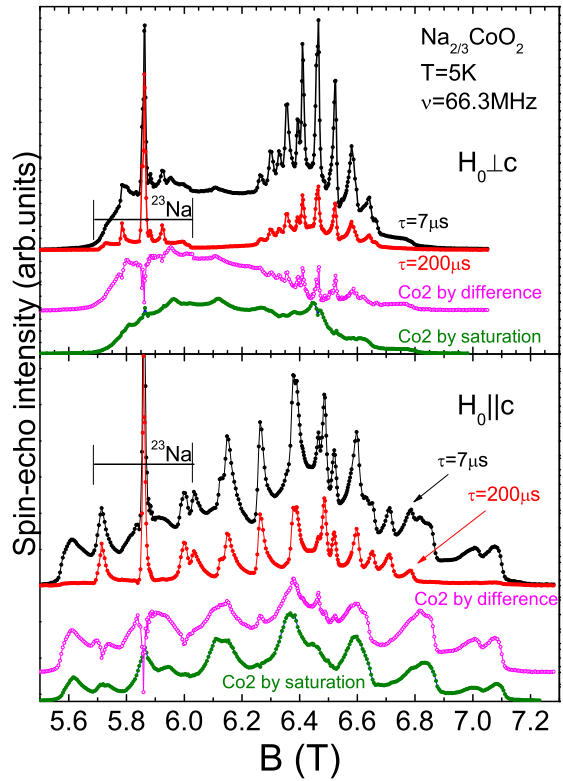}
\caption{(Color online) Examples of $^{59}$Co NMR spectra of an oriented
sample of the Na$_{2/3}$CoO$_{2}$ compound taken with the applied magnetic $%
H_{0}\perp c$ (upper panel) and $H_{0}\parallel c$ (lower panel). In each
panel the top spectra are taken using a short time interval $\protect\tau $
= 7$\protect\mu $s between rf pulses. They involve the NMR signals of all
cobalts in the sample. The signals of the slow relaxing Co1 sites and $^{23}$%
Na are isolated in the second spectra taken with $\protect\tau $ = 200$%
\protect\mu $s. The NMR spectra of the fast relaxing Co2 sites could be
obtained either by subtraction of short-$\protect\tau $ and long-$\protect%
\tau $ spectra (Co2 by difference), or by saturation of the slow relaxing
Co1 and $^{23}$Na NMR signals by an additional rf pulse (Co2 by saturation). Although some remnants of the Na and Co1 signals remain in the difference spectra, the actual quadrupole splitting of the Co2 spectrum is quite well resolved for $H_{0}\parallel c$.}
\label{fig:GrCoExp}
\end{figure}

The raw $^{59}$Co NMR spectra in the Na$_{2/3}$CoO$_{2}$ compound are
complicated - as an example we show in Fig.~\ref{fig:GrCoExp} the NMR
spectra measured in a well oriented sample at $T=$5~K for two directions of
the applied magnetic field $H_{0}$. Former $^{59}$Co NMR data \cite{CoPaper}
taken on this phase have already allowed us to evidence distinct Co NMR
lines, but those NMR spectra were somewhat difficult to analyze fully, as
one needs to determine altogether the EFG parameters and NMR shifts of the
various sites. The experiments were furthermore complicated by the need of a
quasi perfect alignment of the powder sample with respect to the applied
magnetic field.

We found\cite{CoPaper,H67NQRprb} that some Co sites have both very short
nuclear spin-spin ($T_{2}$) and spin-lattice ($T_{1}$) relaxation times - we
denote them as Co2 type. The Co sites which have much longer relaxation
times could be also isolated - we call these cobalts as Co1 type.

In Fig.~\ref{fig:GrCoExp} we show a decomposition of the experimental
spectra in the Co1 and Co2 contributions. The spectrum measured with $\tau
=7\mu s$ between rf pulses contains contributions of the NMR signals of both
type (Co1+Co2) of $^{59}$Co nuclei in the sample, whereas for $\tau =200\mu
s $ the only remaining contribution to the signal is that from the slow
relaxing $^{59}$Co1 nuclei. One can then obtain the spectrum of the
fast-relaxing part Co2 - by subtracting the rescaled slow-relaxing part of
the spectra from the short-$\tau $ spectra. The rescaling can be done
empirically by trial and error, or using an estimation of the $T_2$ decay
for the slow relaxing spectrum. In the Fig.~\ref{fig:GrCoExp} the spectrum
obtained so for the fast-relaxing part Co2 is shown as "Co2 by difference".

Slow and fast-relaxing cobalts can be separated as well by the large
difference in their spin-lattice relaxation times.\cite{H67NQRprb} It is
possible to suppress (or more correctly, to saturate) the signals of $^{23}$%
Na and slow-relaxing Co1 type sites by using an additional $\pi /2$ pulse
with some delay (usually 400-1000~$\mu s$) before the $\pi/2-\tau -\pi /2$
pulse sequence. In the spectrum obtained after such a pulse sequence the
intensities of the slow-relaxing cobalt lines is reduced and the lines of
the fast-relaxing cobalts are better resolved. Spectra of Co2 obtained by
this method are shown in Fig.~\ref{fig:GrCoExp} as "Co2 by saturation".

The shapes of the fast-relaxing Co2 NMR spectra obtained by the two methods
are not exactly the same as one can see in the Fig.~\ref{fig:GrCoExp}. The
compliance between the two methods is much better when the external magnetic
field is applied parallel to the $c$ axis of the sample. The main reason for
the discrepancy is that the nuclear relaxation rates differ in the different
parts of the NMR spectra. Therefore it is not so easy to subtract fully or
to suppress fully the Co1 spectra, so that some remanent of the intense Co1
lines remains in the "Co2 by difference" spectra. Also the overlap with the $%
^{23}$Na NMR signal creates additional difficulties - one can see the
residual traces of the sodium signal in the obtained Co2 spectra.

\subsection{Slow relaxing Co1 NMR spectra}

\begin{figure}[tbp]
\center
\includegraphics[width=1\linewidth]{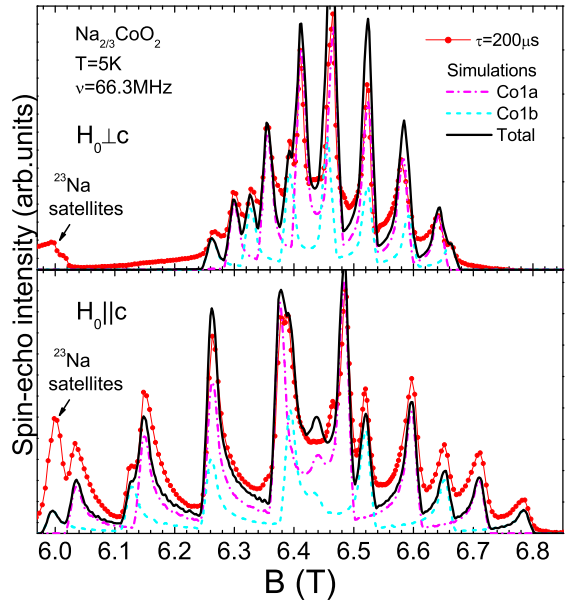}
\caption{(Color online) $^{59}$Co NMR spectra ((red) dots joined by a full
line) of the slow-relaxing Co1 sites measured at $T=$5~K in the two
directions of the applied magnetic field $H_{0}$ which were shown in Fig.~%
\protect\ref{fig:GrCoExp}. The simulation of the spectra of Co1a site are
shown by (magenta) dash-dotted lines, those of the Co1b spectra are shown
by (cyan) dashed lines. The solid black line corresponds to the sum of the
Co1a and Co1b simulated spectra.}
\label{fig:GrCo1Sim5K}
\end{figure}

\begin{figure}[tbp]
\center
\includegraphics[width=1\linewidth]{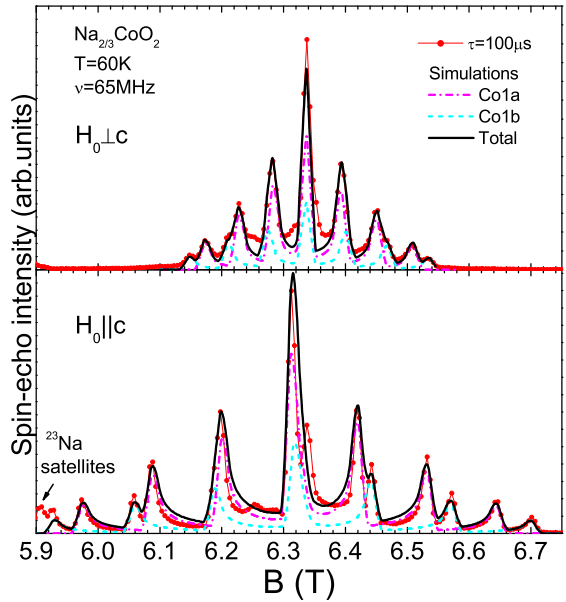}
\caption{(Color online) Experimental $^{59}$Co NMR spectra ((red) dots
joined by a full line) of the slow-relaxing Co1 sites measured at $T=$60~K
in the two directions of the applied magnetic field $H_{0}$. The simulation
of the spectra of Co1a site are shown by (magenta) dash-dotted lines, those
of the Co1b spectra also shown by (cyan) dashed lines. The solid black line
corresponds to the sum of the Co1a and Co1b simulated spectra.}
\label{fig:GrCo1Sim60K}
\end{figure}

Figures~\ref{fig:GrCo1Sim5K} and ~\ref{fig:GrCo1Sim60K} show spectra of the
slow relaxing Co1 at two temperatures, 5~K and 60~K, respectively. These
experimental spectra are quite simple as in both directions of applied
magnetic field $H_{0}\perp c$ and $H_{0}\parallel c$ they are nearly
symmetric with respect to the center of the spectra. The largest splitting
between satellites for slow relaxing Co1 is observed when the applied
magnetic field $H_{0}$ is parallel to the $c$ axis of the sample. This fact
demonstrates that the main principal axis $Z$ of the EFG tensor for Co1 in
the Na$_{2/3}$CoO$_{2}$ compound is parallel to the crystallographic $c$
axis. The fact that the distance between outer singularities in both
directions $H_{0}\perp c$ and $H_{0}\parallel c$ scales as 1:2 shows that
these cobalt sites have an axial EFG tensor with asymmetry parameter $\eta $
close to zero (see Eq.~\ref{eq:SatelDist}).

The $^{59}$Co NMR spectra of the slow relaxing Co1 can be described by only
2 unequivalent cobalt sites which we assign to the Co1a and Co1b sites of Fig.~\ref{fig:StrucCoH67}. In figures ~\ref{fig:GrCo1Sim5K} and ~\ref{fig:GrCo1Sim60K} we show computer simulations of
the $^{59}$Co NMR spectra that we could perform. As one can see the
agreement between the simulated spectra and the data is very good. These
simulations of the NMR spectra allow us to deduce $\nu _{Q}$, $\eta $ and
the magnetic shift $K_{\alpha \alpha}$ of the central line for the two Co1
sites, which are summarized in the Table~\ref{tab:CoSimParam}.

These results are in good agreement with the parameters deduced from $^{59}$%
Co NQR spectra which also display two slow relaxing sites.\cite{H67NQRprb}
Moreover, the simulation of the $^{59}$Co NMR spectra also confirm the ratio
2:1 between the lines intensities corresponding to the Co1a and Co1b sites
as can be seen in Fig.~\ref{fig:GrCo1Sim5K} and ~\ref{fig:GrCo1Sim60K}.

\subsection{Details of the simulations of the NMR spectra}

The simulation procedure used to reproduce the powder NMR spectra requires
an averaging of computed signals for all possible orientations of the powder
particles. In our samples the $c$ axes of the crystallites are aligned but
the $a$ or $b$ axes are at random. To simulate a possible imperfect
alignment of the powder particles, we introduced in our simulation a
distribution of crystallite $c$ axes orientations which we described by a
Lorentzian function. For well oriented samples (like in Fig.~\ref%
{fig:GrCoExp}) the deviation of $c$ axis orientations was within $\pm
5^{\circ }$. Such an angular distribution described very well the $^{23}$Na
NMR spectra (see Appendix~\ref{AppendixPowdAlign}) and has been used then in
the simulations taken for all cobalt sites in a given sample (see figures %
\ref{fig:GrCo1Sim5K}-\ref{fig:GrCo2Sim60K}).

Simulations of the spectra were done taking into account not only the
splittings of the energy levels described by the Hamiltonian of Eq.~\ref%
{eq:Hamilt2} but also the transition probabilities due to the radio
frequency field excitation using the algorithm described in Ref.~%
\onlinecite{Simul}. Also to simulate the broadening of the NMR lines we used
a triangular filtering function with bandwidth 0.15~MHz for both Co1 and Co2
sites at $T$=60~K. At $T$=5~K the larger broadening of the Co2 site spectrum
required an increase of this bandwidth to 0.5~MHz, while 0.15~MHz could be
kept for the Co1 sites.

\begin{table}[tbp]
\caption{Parameters used in Hamiltonian \protect\ref{eq:Hamilt2} for computer simulations of the $^{59}$Co NMR spectra shown in figures \protect
\ref{fig:GrCo1Sim5K}-\protect\ref{fig:GrCo2Sim60K} for different cobalt
sites in Na$_{2/3}$CoO$_{2}$. The principal components of the tensors are
given in the principal axes ($X,Y,Z$) of the EFG tensor. Site occupancy ratios Co1a/Co1b/Co2a/Co2b=1/2/3/6 were used in the simulations.\cite{EPL2009}}
\label{tab:CoSimParam}%
\begin{ruledtabular}
\begin{tabular}{cccccc}
 & & Co1a & Co1b & Co2a & Co2b\\
\hline
& $K_{XX}$ & 1.93 & 2.08 & 9.52 & 10.4 \\
& $K_{YY}$ & 1.93 & 2.08 & 4.36 & 4.76 \\
NMR & $K_{ZZ}$ & 3.48 & 3.20 & 3.54 & 3.80 \\
& $\nu_Q$\footnotemark[1] & 1.19 & 1.386 & 2.276 & 2.591 \\
5~K & $\eta$ & 0 & 0 & 0.362 & 0.358 \\
& $\nu_X$\footnotemark[1] & -0.595 & -0.693 & -0.73 & -0.83 \\
& $\nu_Y$\footnotemark[1] & -0.595 & -0.693 & -1.55 & -1.76 \\
\hline
& $K_{XX}$ & 1.98 & 1.98 & 5.19 & 6.0 \\
NMR & $K_{YY}$ & 1.98 & 1.98 & 2.80 & 3.31 \\
& $K_{ZZ}$ & 2.67 & 2.72 & 2.67 & 3.72 \\
60~K & $\nu_Q$\footnotemark[1] & 1.165 & 1.340 & 2.19 & 2.55 \\
& $\eta$ & 0 & 0 & 0.362 & 0.358 \\
\hline
NQR & $\nu_Q$\cite{H67NQRprb} & 1.193(1) & 1.392(1) & 2.187(1) & 2.541(1) \\
4.2~K & $\eta$\cite{H67NQRprb} & $\leq$0.017 & $\leq$0.016 & 0.362(5) & 0.358(4) \\
\end{tabular}
\end{ruledtabular}
\footnotetext[1]{Since in the NMR the sign of the EFG cannot be determined, the sign of values of $\nu_\alpha$ could be reversed.}
\end{table}

\subsection{Fast relaxing Co2 NMR spectra}

\begin{figure}[tbp]
\center
\includegraphics[width=1\linewidth]{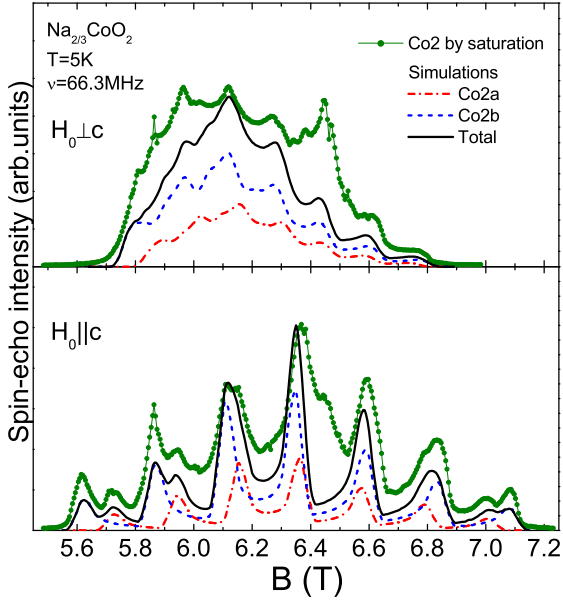}
\caption{(Color online) $^{59}$Co NMR spectra ((green) dots joined by a full
line) of the fast-relaxing Co2 sites measured at $T=$5~K in the two
directions of the applied magnetic field $H_{0}$. Those were obtained by
saturation of slow-relaxing Co1 and $^{23}$Na NMR signals and are taken from
Fig.~\protect\ref{fig:GrCoExp}. The simulation of the spectra of the Co2a
site are shown by (red) dash-dotted lines, those of the Co2b spectra also
shown by (blue) dashed lines. The solid black line corresponds to the sum of
the Co2a and Co2b simulated spectra. The large deviations seen for B$\approx$6.45~Tesla and B$\approx$5.9~Tesla are due to the imperfect saturation of the Co1 and Na contributions.}
\label{fig:GrCo2Sim5K}
\end{figure}

\begin{figure}[tbp]
\center
\includegraphics[width=1\linewidth]{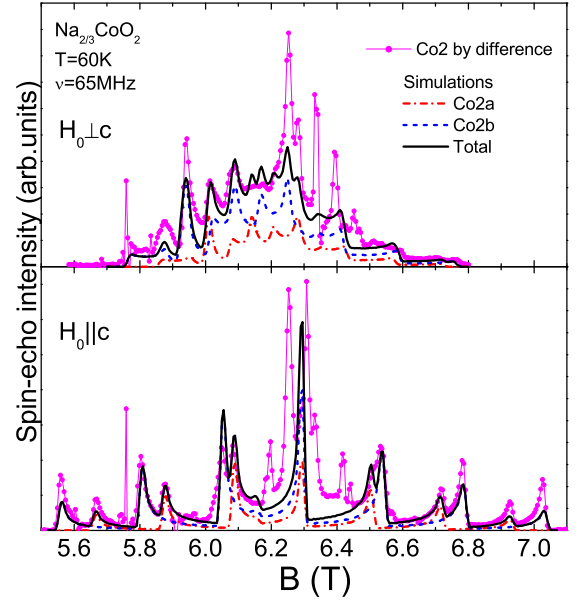}
\caption{(Color online) $^{59}$Co NMR spectra ((magenta) dots joined by a
full line) of the fast-relaxing Co2 sites measured at $T=$60~K in the two
directions of the applied magnetic field $H_{0}$ . Those were obtained by
subtraction of long-$\protect\tau $ spectra from the short-$\protect\tau $
spectra. The simulation of the spectra of the Co2a site are shown by (red)
dash-dotted lines, those of the Co2b spectra also shown by (blue) dashed
lines. The solid black line corresponds to the sum of the Co2a and Co2b
simulated spectra. The large peaks between 6.2 and 6.4~Tesla are remnants of the Co1 NMR signals, incompletely subtracted.}
\label{fig:GrCo2Sim60K}
\end{figure}

Figures~\ref{fig:GrCo2Sim5K} and ~\ref{fig:GrCo2Sim60K} show the spectra of
the fast relaxing Co2 at two temperatures, 5~K and 60~K, respectively.
Despite the difficulties in obtaining fast relaxing Co2 NMR spectra, some
facts can be settled. First of all the largest splitting between satellites
for Co2 is observed for $H_{0} \parallel c$. Therefore for Co2 the main
principal axis $Z$ of the EFG tensor is parallel to the crystallographic $c$
axis in the Na$_{2/3}$CoO$_{2}$ compound.

The Co2 spectrum for $H_{0}\parallel c$ at $T=$60~K (see Fig.~\ref%
{fig:GrCo2Sim60K}) is quite simple, well resolved and clearly evidences two
inequivalent Co2 sites with different values of quadrupole frequency $\nu
_{Q}$ - we denote these sites as Co2a and Co2b. The relative intensities of
the NMR signals of these sites corresponds to a ratio Co2a:Co2b close to
1:2. These facts are in very good agreement with the $^{59}$Co NQR spectra
which also display two fast relaxing sites with a similar intensity ratio.%
\cite{H67NQRprb}

At low temperature $T=$5~K the Co2 spectrum in the $H_{0}\parallel c$ case
(see Fig.~\ref{fig:GrCo2Sim5K}) is less resolved due to the large magnetic
broadening of the lines, but the satellites of the two sites could still be
seen.

If we compare now the Co1 and Co2 NMR spectra measured in the $H_{0}\perp c$
it is easy to see that Co2 spectra are asymmetric with a larger intensity in
the lower fields. At $T$=60~K some lines are resolved with clear
singularities but at $T=$5~K it becomes nearly impossible to resolve the
different lines and we could only detect very weak singularities. As the
orientation procedure of our samples is such that the $ab$ planes of
different crystallites are randomly distributed then the experimental
spectra in the $H_{0}\perp c$ direction are powder spectra. This explains
why they are less resolved than the $H_{0}\parallel c$ spectra. However some
singularities could still be perceived even in the broadened spectrum at $T=$%
5~K (Fig.~\ref{fig:GrCo2Sim5K}). Such a complexity is apparent as well in
the data taken by many others groups in sodium cobaltates with sodium
content $x\approx 0.7$.\cite{Ishida07, GavilanoCo07} Comparing our oriented
powder NMR spectra with NMR data taken on single crystals \cite%
{ImaiPRL1,MHJulien075} allows us to conclude that twinning of the $ab$ plane
is quite common in those single crystals.

Therefore the analysis of a single $H_{0}\perp c$ spectrum of Co2 appears
quite difficult. Independently of the method used to isolate the fast
relaxing Co2 spectra we do however find that the positions of the different
singularities in the Co2 spectra are the same as one can see in Fig.~\ref%
{fig:GrCoExp}. As usual such singularities in a powder NMR spectrum appear
at positions where the two principal axes $X$ and $Y$ of the EFG and
magnetic shift tensor in the $ab$ plane correspond to extremal positions of
the NMR signal for randomly distributed crystallites. Furthermore, as was
shown in Ref.~\onlinecite{CoPaper} the sodium and cobalt sites in the Na$%
_{2/3}$CoO$_{2}$ compound feels a common $T$ dependence of the spin
susceptibility, which we could determine directly by measuring the $^{23}$Na
magnetic shift $^{23}K(T)$.\cite{NaPaper} We did then study the evolution of
the positions of the singularities in the Co2 spectrum versus temperature.
By plotting these singularities versus the spin susceptibility as monitored
by $^{23}K$, the temperature being then an implicit parameter, we could sort
out those singularities which correspond to Co2 central lines and those
which can be assigned to quadrupolar satellites, as detailed in Appendix~\ref%
{AppendixCo2XY}.

This allowed us to resolve then fully the EFG and shift tensors of Co2a and
Co2b sites and to understand that not only the EFG tensor but also the shift
tensors are highly asymmetric. We could then succeed to perform a full
simulation of the $^{59}$Co NMR spectra of the fast relaxing cobalts in the
two field directions, as shown in figures ~\ref{fig:GrCo2Sim5K} and ~\ref%
{fig:GrCo2Sim60K}, with the parameters collected in the Table~\ref%
{tab:CoSimParam}. There we also reported the parameters measured for the Co1a
and Co1b sites, as obtained from the simulations done in Fig.~\ref%
{fig:GrCo1Sim5K} and Fig.~\ref{fig:GrCo1Sim60K}. As one can see the obtained
values of the quadrupole frequency $\nu _{Q}$ and of the asymmetry parameter
$\eta $ for all sites are in good agreement with the values obtained by NQR.%
\cite{H67NQRprb}

\section{$T$ dependence of the cobalt magnetic shifts}

\label{TdepCoshift}

The principal components of the magnetic shift tensor for the cobalt site $i$
in a direction $\alpha $ consist on three parts
\begin{equation}
\begin{array}{l}
K_{i,\alpha }=K_{i,\alpha }^{dia}+K_{i,\alpha }^{orb}+K_{i,\alpha }^{spin}
\\
\\
\;\;\;\;\;\;\;\;=K_{i,\alpha }^{dia}+\ A_{i,\alpha }^{orb}\;\chi _{i,\alpha
}^{orb}+\;A_{i,\alpha }^{s}\;\chi _{i,\alpha }^{s}(T).  \label{eq:Kcomp}
\end{array}%
\end{equation}%
Here $K_{i,\alpha }^{dia}$ is the chemical shift due to the diamagnetic
susceptibility of the inner shell electrons, $K_{i,\alpha }^{\mathrm{orb}}$
reflects the on site orbital effect of the valence electrons, proportional to the orbital part of the electronic susceptibility $\chi _{i,\alpha }^{orb}$ which is usually temperature independent. The last term, the spin shift $K_{i,\alpha }^{\mathrm{spin}}$, is proportional to the local
electronic spin susceptibility $\chi_{i,\alpha }^{s}(T)$.

The remarkable feature of the Na$_{2/3}$CoO$_{2}$ phase of the sodium
cobaltates is the large low $T$ variation of this last term, which, as already pointed out above, allowed us to resolve the Co NMR spectra.\cite{EPL2008} The $^{23}$Na NMR has negligible chemical and orbital NMR shifts and is largely dominated by the $T$ dependent spin term which has been found nearly isotropic.\cite{NaPaper} So the comparison of $^{59}K_{i,\alpha }$ with the isotropic contribution $^{23}K_{iso}$ also allowed us to separate the orbital and spin contributions to the cobalt magnetic shift, as detailed below.

\subsection{Co1 shifts}

In Fig.~\ref{fig:GrCo1Shift}a we show the $T$ dependence of the magnetic
shifts of Co1a and Co1b sites which are plotted in Fig.~\ref{fig:GrCo1Shift}%
b versus the NMR shift of $^{23}$Na. For Co1a and Co1b sites the $Z$%
-components of the magnetic shift are quite similar, with a substantial $T$
dependence which is proportional to $^{23}K$ at temperatures below 150~K.

The in-plane $X$ and $Y$ components of the shift are indistinguishable for
Co1a and Co1b sites, and their shift value does not change from 150~K down to
$T=$40~K. Below this temperature some small difference in the in-plane
shifts for Co1a and Co1b sites appears. This was also seen
by others, \cite{GavilanoCo07} which points out that this effect is somewhat
characteristic of this $x=2/3$ phase of Na cobaltates. But as one can see in
Fig.~\ref{fig:GrCo1Shift}, these distinct variations of NMR shifts of Co1a
and Co1b are quite small when compared with the $T$ variations of the NMR
shifts of both Co1 sites for $H_{0}\parallel c$. So we can consider so far
that the in plane shift of the Co1 sites is practically $T$ independent
below 150~K.

Using Fig.~\ref{fig:GrCo1Shift}b one can get the slopes of the linear fits
of $^{59}K$ versus $^{23}K$ which give the relative magnitudes of the spin
contributions $K_{i,\alpha }^{\text{s}}/^{23}K$, while the $^{23}K$=0
intercepts give the estimates of the $T$ independent contribution to the $%
^{59}$Co NMR shift. The values found here, reported in Table~\ref%
{tab:CoShifts}, are much larger than those expected for the chemical shift $%
K_{i,\alpha }^{dia}$ $\lesssim 0.1\%$. They are obviously dominated here by the Co orbital NMR shift $K_{i,\alpha }^{orb}$, so that we have neglected $K_{i,\alpha}^{dia}$ in all our analyzes.

But, as seen in Fig.~\ref{fig:GrCo1Shift}, significant increases of $%
^{59}K_{1}^{\alpha }$ with respect to the low $T$ linear $^{23}K$ dependence
are observed for $T\gtrsim 150$~K, that is for $^{23}K<0.08$. Those
increases are associated with the onset of Na motion which has been detected
at $\approx $200~K from $^{23}$Na data in Ref.~\onlinecite{EPL2008}. There we have discussed that this Na motion induces a significant Co1-Co2 site exchange
which induces such an increase of Co1 NMR shifts, while the charge
disproportionation already occurs above room temperature.

\subsection{Co2 shifts} \label{Co2shifts}

In Fig.~\ref{fig:GrCo2Shift}a we show the $T$ dependencies of the magnetic
shifts of Co2a and Co2b sites and in Fig.~\ref{fig:GrCo2Shift}b they are
shown versus $^{23}K$. In this figure we put some experimental
points which were determined accurately at specific temperatures, but due to
the relatively small number of Co2a sites the intensity of its NMR signal is
weak. For this site we plot lines corresponding to the $K_{X}$ and $K_{Y}$
components of its shift obtained from the analysis detailed in Appendix B.

\begin{figure}[tbp]
\center
\includegraphics[width=1\linewidth]{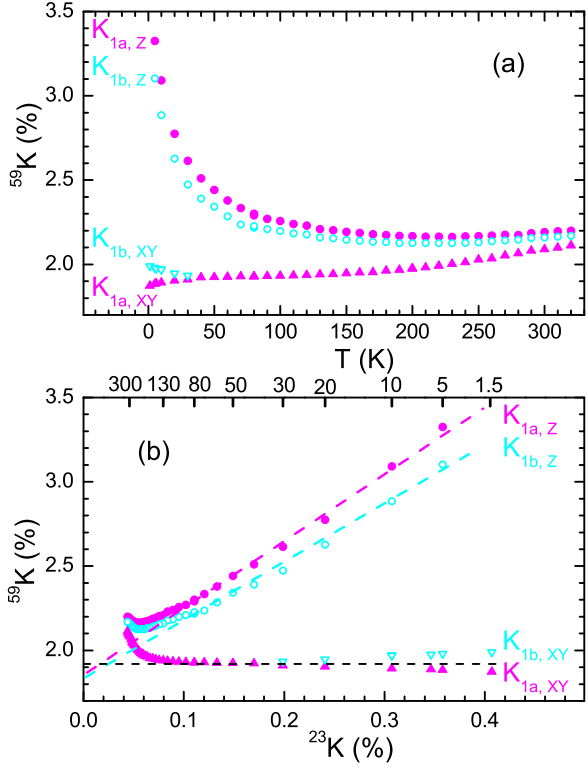}
\caption{(Color online) (a) $T$-dependencies of the $^{59}$Co NMR shifts of
the Co1a (filled (magenta) symbols) and Co1b (open (cyan) symbols) sites for
two directions of the external magnetic field $H_{0}$. For $H_{0}\parallel Z$
(circles) the Co1a and Co1b shifts exhibit the same $T$ dependence with a
small difference in absolute values. In the $H_{0}\perp Z$ direction for
both sites Co1a and Co1b the $T$ dependence is much weaker and a small
difference between the shifts of these sites appears only below $T$=40~K.
For both Co1a and Co1b sites the magnetic shifts are axial. (b) The data of
(a) are plotted versus the $^{23}$Na NMR shift showing linear slopes at low
temperatures and deviations from this linear dependence above 100~K. Such
low-$T$ linearity allowed us to determine the orbital $K_{ii}^{\text{orb}}$
and spin $K_{ii}^{\text{s}}$ contributions to $^{59}K$ for all three
directions $i=X,Y,Z$ - see Table~\protect\ref{tab:CoShifts}.}
\label{fig:GrCo1Shift}
\end{figure}

\begin{figure}[tbp]
\center
\includegraphics[width=1\linewidth]{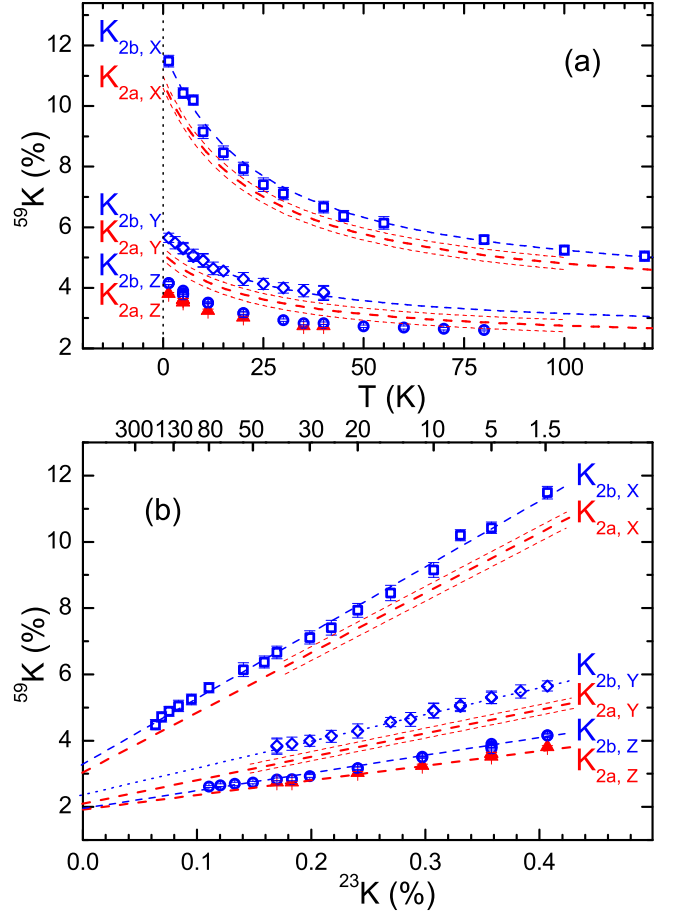}
\caption{(Color online) (a) $T$-dependencies of the $^{59}$Co NMR shifts of the Co2a ((red) closed symbols and (red) dashed lines) and Co2b ((blue) open
symbols and (blue) dashed lines) sites for the three directions of the
external magnetic field $H_{0}$ parallel to the $X,Y,Z$ principal axes of
the EFG tensor. The method used to determine the data for $H_{0}\parallel
X,Y $ directions are detailed in Appendix~\protect\ref{AppendixCo2XY}. (b)
The data of (a) are plotted versus the $^{23}$Na NMR shift. The linear
variation of $^{59}K$ versus $^{23}K$ allowed us to determine the orbital $%
K_{ii}^{\text{orb}}$ and spin $K_{ii}^{\text{s}}$ contributions to the $%
^{59}K$ for the three directions $i=X,Y,Z$ - see Table~\protect\ref%
{tab:CoShifts}.}
\label{fig:GrCo2Shift}
\end{figure}

As one can see in Fig.~\ref{fig:GrCo2Shift}b a linear dependence is found
for all components of the magnetic shift for both Co2a and Co2b sites below
120~K in a range where no sign of Na motion could be seen. So this figure in combination with Fig.~\ref{fig:GrCo1Shift}b confirm that all 4 cobalt sites pertain indeed to a unique sodium cobaltate phase in which a single $T$ variation characterizes the local $\chi _{i,\alpha }^{s}(T)$. We collected in Table~\ref{tab:CoShifts} the relative magnitude of the spin contributions $K_{i,\alpha }^{\text{s}}/^{23}K$ and $K_{i,\alpha }^{orb}$ for all three components of the magnetic shift for the Co2a and Co2b sites together with those of the Co1 sites.

While the magnetic shift tensors are axial for the Co1a and Co1b sites, we
see in Table~\ref{tab:CoShifts} that both Co2a and Co2b sites demonstrate
strong in-plane anisotropy of both the orbital shift and the spin hyperfine
coupling. \textit{This totally unexpected anisotropy is a quite important
observation achieved in the present investigation.} In our previous paper
concerning the $^{59}$Co NMR in this phase \cite{CoPaper} as well as in all
the existing publications \cite{Ishida07,GavilanoCo07,MHJulien075,ImaiPRL1}
such anisotropy was not anticipated. This had consequently led us to invoke
a third group of axial cobalt sites Co3 with large shift values. In appendix %
\ref{AppendixPowdAlign} we detail why imperfect alignment of the sample
powder had led us as well to overlook initially this anisotropy.

\begin{table}[tbp]
\caption{ $^{59}$Co magnetic shift parameters obtained for the 4 cobalt
sites.}
\label{tab:CoShifts}%
\begin{ruledtabular}
\begin{tabular}{ccccc}
Site & Co1a & Co1b & Co2a & Co2b\\
\hline
${K_{XX}^{\text{s}}/^{23}K}$ & -0.2(2)  & 0.2(2)  & 17.9(4) & 19.9(4)\\
${K_{YY}^{\text{s}}/^{23}K}$ & -0.2(2)  & 0.2(2)  & 7.1(3)  & 7.9(3) \\
${K_{ZZ}^{\text{s}}/^{23}K}$ & 4.0(2)   & 3.5(2)  & 4.6(2)  & 5.2(2) \\
\hline
${K_{XX}^{\text{orb}}}$(\%)  & 1.92(2)  & 1.92(2) & 3.1(2)  & 3.3(2) \\
${K_{YY}^{\text{orb}}}$(\%)  & 1.92(2)  & 1.92(2) & 2.1(2)  & 2.3(2) \\
${K_{ZZ}^{\text{orb}}}$(\%)  & 1.86(5)  & 1.86(5) & 1.92(5) & 1.95(4)\\
\end{tabular}
\end{ruledtabular}
\end{table}

\section{$^{59}$C\lowercase{o} spin lattice relaxation}

\label{SectionCoT1}

When thermal equilibrium of the nuclear spins is disturbed by rf pulses, the
equilibrium nuclear magnetization is recovered by various relaxation
processes which reflect the interactions in the spin system and the magnetic
and electronic properties of the materials. The recovery of the nuclear
magnetization along the applied field operates through nuclear spin lattice
relaxation (NSLR) processes characterized by the spin-lattice relaxation
time $T_{1}$ or by the nuclear spin-lattice relaxation rate $1/T_{1}$.

In systems with unpaired spins the dominant $T_{1}$ process is due to local
field fluctuations induced by the dynamics of the local electronic
magnetization. Theoretically, the spin contributions to $(T_{1}T)^{-1}$ may
be written using the imaginary part of the Co dynamical electron
spin-susceptibility ${\chi ^{\prime \prime }(\mathbf{q},\nu _{n})}$ as
\begin{equation}
(T_{1}T)^{-1}=\frac{2\gamma _{n}^{2}k_{B}}{g^{2}\mu _{B}^{2}}\sum_{\mathbf{q}%
}|A(\mathbf{q})|^{2}\frac{\chi ^{\prime \prime }(\mathbf{q},\nu _{n})}{\nu
_{n}},  \label{eq:1/T1T}
\end{equation}%
where $A(\mathbf{q})=A_{0}+\sum A_{i}e^{i\mathbf{q}\mathbf{r}_{i}}$ is the
wave-vector \textbf{q} dependent hyperfine form factor which depends on the
hyperfine interaction between the observed Co nuclear and electron spins at
the same site $A_{0}$ and the hyperfine interaction with electron spins at
nearest neighbor Co sites $A_{i}$.\cite{Moriya}

Our measurements of the $^{23}$Na NSLR temperature dependence allowed us to
establish that in sodium cobaltates with sodium content $x>0.6$ the dominant
correlations in the electron spin system are ferromagnetic, while for lower
sodium contents antiferromagnetic correlations take over.\cite{EPL2008,LangNFD}

Equation~\ref{eq:1/T1T} applied to different nuclei tells us that NSLR data
should be similar for all nuclei in the absence of any cancelation at some
specific $\mathbf{q}$ vector due to the form factor associated with the
geometrical position of the nuclear site with respect to the magnetic atoms.
So the comparison of $^{59}$Co and $^{23}$Na NSLR could help us to better
characterize the dynamic susceptibility. We report then here the results of
a detailed study of the $^{59}$Co NSLR in this Na$_{2/3}$CoO$_{2}$ phase.
First we demonstrate the consistency and validity of our measurements, and
then compare the results for the different Co sites. We follow as well the
procedure used for the shift measurements and compare directly the Co and Na
$T_{1}$ data.

\subsection{Experimental techniques}

To study the nuclear spin-lattice relaxation process of the $^{59}$Co we
have used the usual magnetization inversion recovery method with three
pulses: $\pi -t-\pi /2-\tau -\pi$. In this sequence the first pulse rotates
the magnetization by 180$^o$ and the longitudinal magnetization recovered
after time $t$ is measured with a spin echo sequence with interval $\tau$.
The dependence of the spin-echo intensity on delay time $t$ allows to
monitor the recovery of the nuclear magnetization associated with a given
NMR transition:
\begin{equation}
M(t)=M_{0}(1-B\cdot R(t)).  \label{eq:Kinetics}
\end{equation}
Here $M_{0}$ is the thermal equilibrium value of magnetization and the
parameter $B$ characterizes the actual magnetization after the first pulse
at $t=0$ (the imperfection of the experimental conditions gave typical
values $B$ $\simeq $ 1.8 rather than $B=2$ expected for a perfect $\pi $
pulse).

The shape of the relaxation function $R(t)$ depends on the nuclear
transition sampled. For a two-level nuclear system (like the $I=3/2$ NQR
case) this process is exponential $R(t)=exp\left( -t/T_{1}\right) $ and
allows a simple experimental determination of $T_{1}$.\cite{Slichter} But
generally for $I>1/2$ the nuclear energy levels are differentiated by the
quadrupole interaction with the crystalline electric field (Eq.~\ref%
{eq:Hamiltonian}). For very broad NMR spectra as those considered here, the
applied first pulse only inverts the populations of some levels which, for a
given frequency, can be selected by properly choosing the applied field.
Consequently the difference in population between adjacent levels which are
probed by the rf pulses depends on the populations of the levels which are
not hit by the rf pulses. Therefore the magnetization recovery becomes
multi-exponential
\begin{equation}
R(t)=\sum_{i}a_{i}exp\left( -\frac{\lambda _{i}t}{T_{1}}\right) ,
\label{eq:Recovery}
\end{equation}%
but is still characterized by a single $T_{1}$ value.

The spin-lattice relaxation could be driven either by magnetic or
quadrupolar fluctuations which correspond to distinct transition
probabilities and $\lambda _{i}$ values. However in the sodium cobaltates
the magnetic relaxation mechanism dominates at least at low $T$.\cite%
{H67NQRprb} Using Ref.~\onlinecite{Andrew,McDowell} we have calculated for
our values of $\eta$ the $a_{i}$ and $\lambda_{i}$ parameters for the
theoretical relaxation functions $R(t)$ for the case of magnetic relaxation
by weak fluctuating magnetic fields for the different $^{59}$Co transitions
- see Table~\ref{tab:CoT1Coef}.

\begin{table}[tbp]
\caption{Parameters $\protect\lambda_{i}$ and $a_{i}$ of the theoretical
spin-lattice relaxation function (\protect\ref{eq:Recovery}) for different
transitions of $^{59}$Co for the case of magnetic relaxation by weak
fluctuating magnetic fields (based on Ref.~\onlinecite{Andrew,McDowell}).}
\label{tab:CoT1Coef}%
\begin{ruledtabular}
\begin{tabular}{cccccccc}
$\lambda_i$ & 28 & 21 & 15 & 10 & 6 & 3 & 1 \\
Transition & $a_1$ & $a_2$ & $a_3$ & $a_4$ & $a_5$ & $a_6$ & $a_7$ \\
\hline
$-\frac{1}{2}\leftrightarrow +\frac{1}{2}$ & 0.714 & 0 & 0.206 & 0 & 0.068 & 0 & 0.012 \\
$\pm \frac{1}{2}\leftrightarrow \pm \frac{3}{2}$ & 0.457 & 0.371 & 0.001 & 0.117 & 0.030 & 0.012 & 0.012\\
$\pm \frac{3}{2}\leftrightarrow \pm \frac{5}{2}$ & 0.114 & 0.371 & 0.366 & 0.081 & 0.008 & 0.048 & 0.012\\
$\pm \frac{5}{2}\leftrightarrow \pm \frac{7}{2}$ & 0.009 & 0.068 & 0.206 & 0.325 & 0.273 & 0.107 & 0.012\\
\end{tabular}
\end{ruledtabular}
\end{table}

\subsection{$^{59}$Co spin lattice relaxation results}

We have already seen by NQR that $T_{1}$ data are identical for the Co1a and
Co1b sites. They are identical as well for Co2a and Co2b sites but much
shorter than on the Co1 sites, which allowed us to confirm that the Co2
sites were indeed the magnetic sites.\cite{H67NQRprb}

\begin{figure}[tbp]
\center
\includegraphics[width=1\linewidth]{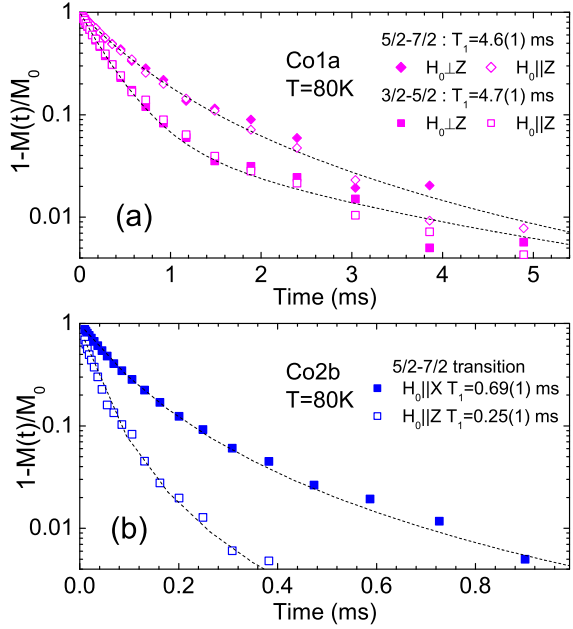}
\caption{(Color online) Spin-lattice relaxation curves measured at $T=80$~K
for two orientations of the magnetic field. (a) for Co1a sites the data
shown have been taken on two satellite transitions $\frac{5}{2}
\leftrightarrow \frac{7}{2}$ and $\frac{3}{2}\leftrightarrow \frac{5}{2}$.
The fits of the data with the magnetization relaxation functions (\protect
\ref{eq:Recovery}), with parameters from the Table~\protect\ref{tab:CoT1Coef}%
, give a unique $T_{1}$ value, independent of the field orientation. (b) For
Co2b similar good fits are obtained for the data, but the relaxation time
depends on the field orientation.}
\label{fig:GrCoT1Curves}
\end{figure}

Due to the close values of the magnetic shift for all 4 cobalt sites in the $%
Z$ direction (see Table~\ref{tab:CoShifts}), to avoid cross-relaxation
between sites it is of course better to take data on quadrupolar satellite
transitions of different sites which are quite separated in frequency (or
field). So we have measured the spin-lattice relaxation curves on the outer
satellites $+\frac{5}{2}\leftrightarrow +\frac{7}{2}$ for the sites with
larger signal intensity, that is Co1a for the slow relaxing sites and Co2b
for the fast relaxing sites. In Fig.~\ref{fig:GrCoT1Curves}, we display such
examples of experimental spin-lattice relaxation curves for Co1a and Co2b
sites. One can see there that they can be fitted quite well by the function %
\ref{eq:Recovery} with the coefficients from Table~\ref{tab:CoT1Coef}. It
can also be seen in Figure~\ref{fig:GrCoT1Curves}a that the same $T_{1}$
value is deduced from the data for the $\frac{3}{2}\leftrightarrow \frac{5}{2%
}$ and $\frac{5}{2}\leftrightarrow \frac{7}{2}$ transitions for the Co1a
nucleus, demonstrating convincingly that the data analysis is reliable and
that the relaxation is indeed magnetic.

We could also evidence in Fig.~\ref{fig:GrCoT1Curves}b that the $T_{1}$ data depend of the orientation of the applied field for Co2, contrary to the isotropic results found for the Co1 sites in Fig.~\ref{fig:GrCoT1Curves}a.

\begin{figure}[tbp]
\center
\includegraphics[width=1\linewidth]{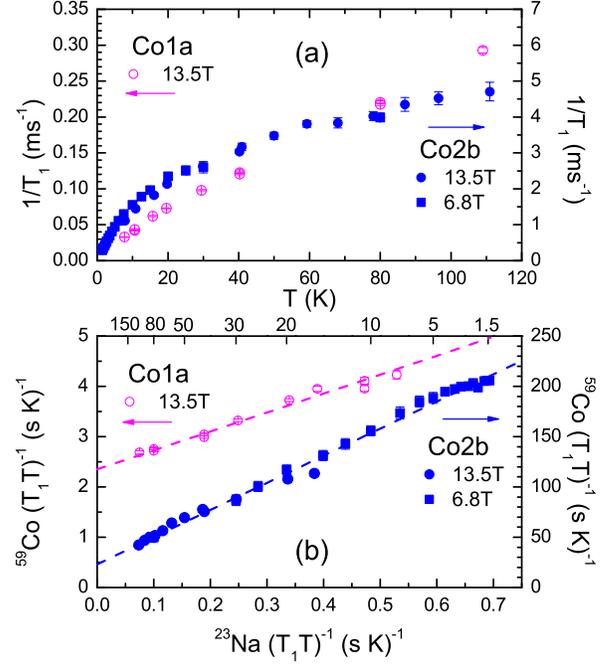}
\caption{(Color online) (a) Temperature dependence of the NSLR rate $T_{1}^{-1}$ measured in $H_0 \parallel c$ orientation for the Co1a (magenta empty circles) and Co2b sites (blue closed
circles and squares, taken for two different applied fields). (b) The $%
(T_{1}T)^{-1}$ of $^{59}$Co of (a) are plotted versus $(T_{1}T)^{-1}$ values
of $^{23}$Na, the temperature $T$ being then an implicit parameter in the
plot. For both Co1a and Co2b sites the variations are found linear, which
allows us to separate the spin and orbital contributions to the $^{59}$Co $%
(T_{1}T)^{-1}$(see text).}
\label{fig:GrCoT1}
\end{figure}

Similarly, in Fig.~\ref{fig:GrCoT1}a we plotted then the $T$ variation of
the Co2b $T_{1}^{-1}$ data measured in two distinct applied fields, showing that $T_{1}$ values measured are identical, which confirms the reliability of the measurements, in view of the large spectral changes which occur with
increasing field. We also compare there the data taken on Co1a with those on
Co2b. The T$_{1}$ values differ by a factor as large as 20 which confirms
the strong differentiation already found in previous observations done by
NQR at low $T$.\cite{H67NQRprb}

Generally, as given in Eq.~\ref{eq:1/T1T} the electron spin contribution to the spin-lattice relaxation rate $T_{1}^{-1}$ is governed by the fluctuations of
the transverse components of the effective magnetic field induced at the
nucleus by the electronic magnetization. In NMR the quantization axis is
given by the external magnetic field and for the $^{59}$Co $T_{1}^{-1}$ the
dominant term is the fluctuating hyperfine field whose magnitude is
proportional to the hyperfine field $|A(\mathbf{q})|$ and local spin
susceptibility ${\chi ^{\prime \prime }(\mathbf{q},\nu _{n})}$. The NMR
shift and $T_{1}$ data taken on $^{23}$Na have given evidence that these electronic susceptibilities are nearly isotropic,\cite{NaPaper,EPL2008} so the spin contributions to $1/T_{1}$ on all nuclei are expected to scale with each other and the $T_{1}$ anisotropies should be governed by those of the hyperfine couplings $|A(\mathbf{q})|$. For instance
\[
T_{1Z}^{-1}\propto (K_{YY}^{2}+K_{XX}^{2})\ \text{and }T_{1X}^{-1}\propto
(K_{YY}^{2}+K_{ZZ}^{2})
\]%
Therefore as the spin component of the magnetic shift $K_{XX}$ for Co2 is
much larger than that for Co1 (see Table~\ref{tab:CoShifts}) it is quite
natural to find much shorter $T_{1}$ values for Co2 than for Co1. Also, as
the components of the magnetic shift tensors for Co2 sites are highly
anisotropic, with a dominant value for $H_{0}\parallel X$, the transverse
fluctuating local fields are expected much larger for $T_{1Z}^{-1}$ than for
$T_{1X}^{-1}$ in qualitative agreement with the anisotropy of the $T_{1}$  data displayed in Fig.~\ref{fig:GrCoT1Curves}b.

However, for Co1 a distinct situation occurs. As one can see in Fig.~\ref%
{fig:GrCoT1Curves}a for Co1a the spin-lattice relaxation is isotropic while
the spin component to the shift is strongly anisotropic - $K_{ZZ}$ is much
larger than the in plane components which are vanishingly small - see Fig.~%
\ref{fig:GrCo1Shift} and Table~\ref{tab:CoShifts}.

To explain that, we are lead to consider whether the spin components of the
electronic dynamic susceptibility do dominate the relaxation, or is there
another independent contribution to $1/T_{1}$, such as for instance an orbital  contribution to the spin lattice relaxation? Indeed, usually for transition metals two additive contributions to $T_{1}$ are expected
\begin{equation}
\frac{1}{TT_{1}}=\frac{1}{TT_{1}^{orb}}+\frac{1}{TT_{1}^{spin}}.
\label{eq:T1comp}
\end{equation}%
So we tried to separate such components similarly to what we had done
before for the cobalt shifts - see Section~\ref{TdepCoshift}. Indeed, as was
shown in Ref.~\onlinecite{EPL2008} in the Na$_{2/3}$CoO$_{2}$ compound the $%
^{23}$Na spin-lattice relaxation is largely dominated by the $T$ dependent
spin term while any orbital contribution is negligible on that nuclear site.
So we can plot the values of $(T_{1}T)^{-1}$ of $^{59}$Co versus $%
(T_{1}T)^{-1}$ values of $^{23}$Na, as done in Fig.~\ref{fig:GrCoT1}b. Good
linear relations are found for both sites, that is
\[
^{59}(T_{1}T)_{Z}^{-1}\ =\ ^{23}(T_{1}T)^{-1}\ast 3.8(1)+2.4(1)
\]
for Co1a and
\[
^{59}(T_{1}T)_{Z}^{-1}\ =\ ^{23}(T_{1}T)^{-1}\ast 270(3)+23(2)
\]
for Co2b. This analysis allows indeed to separate in the $T_{1}$ data the
spin part from an orbital $T_{1}T$ = constant part.\cite{Obata} Let us
notice that the ratio of the spin contributions for Co2 and Co1 is about $%
270/3.8 \approx 70$. Such a large ratio compatible with that obtained by NQR%
\cite{H67NQRprb} is quite expected as the spin part of the Co1 shift in the $%
X$ and $Y$ directions is practically zero - see Table~\ref{tab:CoShifts}.

As for the orbital contribution, one can see in Fig.~\ref{fig:GrCoT1} that
it is small with respect to the spin contribution for the Co2 site while it
becomes dominant for the Co1 site. For instance at $T$=80~K the $T_{1}$ is
totally dominated by the orbital term for the Co1 site, which perfectly
justifies the absence of detected $T_{1}$ anisotropy on the Co1 site at this
temperature. Let us point out however that the orbital term is still one
order of magnitude larger for Co2 than for Co1.

Overall these $T_{1}$ data are totally compatible with the anisotropy of the
Co2 shifts reported in Sec.~\ref{Co2shifts}. We shall discuss further the implication of these
results on the electronic properties of this Na$_{2/3}$CoO$_{2}$ phase in
the following discussion section.

\section{Discussion}

\label{SectionDiscussion}

We have reached now a rather complete experimental determination of the NMR
parameters for the four Co sites formerly detected by NQR,\cite{H67NQRprb}
which, together with the $^{23}$Na NMR spectra allowed us to determine the
atomic structure of this Na$_{2/3}$CoO$_{2}$ phase.\cite{EPL2009} The
analysis of the NMR shifts and of the magnitude of the EFGs will allow us to
confirm here the importance of cobalt charge disproportionation between the
non-magnetic and magnetic sites, that is of the charge order induced in the
structure. We shall discuss hereafter how this electronic kagom\'{e}
differentiation is reflected on the orbitals involved at the Fermi level at
the different sites of the Co plane. The anisotropy of the EFG and NMR spin
shifts will be seen to be related with the hole orbital ordering which
governs the electronic properties of this phase.

\subsection{Site differentiation}

The data of Table~\ref{tab:CoSimParam} allowed us to evidence that for the
Co1a and Co1b sites the symmetry of the magnetic shift tensor $\hat{K}$ is
axial ($K_{X}$=$K_{Y}$), as found before for the EFG tensor ($\eta =0$ or $%
\nu _{X}$=$\nu _{Y}$). The local symmetry of these sites had been a key
argument used to resolve the crystal structure.\cite{EPL2009} We also fully
confirm here that the electronic properties are quasi identical for Co1a and
Co1b sites which are slow relaxing and weakly sensitive to the magnetism.

The knowledge of this actual atomic structure has allowed us also to resolve
fully the difficulties connected with the imperfect alignment of the powder
grains for some samples, which had initially lead us to conclude that the
two other Co sites displayed quite distinct nearly axial NMR shifts.\cite%
{CoPaper} The analysis of the extensive set of $^{59}$Co NMR spectra done
for the two fast relaxing sites Co2a and Co2b allow us to
establish that these sites have nearly identical spin shifts tensors which
are unexpectedly totally non-axial, the larger NMR spin shifts being in a
privileged $X$ direction in plane. The large anisotropy of the spin contribution to the $1/T_{1}$ data agrees perfectly with this NMR shift anisotropy, as $T_{1Z}$ is found shorter than $T_{1X}$.

We therefore established here that these fast relaxing Co2a and Co2b sites
have similar anisotropic orbital and spin shifts and almost identical $T_{1}$
values,\cite{EPL2009,H67NQRprb} confirming that they contribute identically
to the electronic properties and are dominantly responsible for the magnetic
properties of this compound.

This allows us then to conclude that although the local environment of Na$%
^{+}$ charges shown in Fig.~\ref{fig:StrucCoH67} are quite distinct for the
two Co1 sites (or the two Co2 sites), the electronic and magnetic properties
only marginally differ within each group of two sites. This establishes
that, when dealing with the electronic properties, one might only consider
\textit{a two site structure, with Co2 on a kagom\'{e} planar structure, and
Co1 on the complementary triangular structure}.

\subsection{Charge ordering}

This differentiation of sites had clearly led us to suggest that the ionic
characters of the Co1 and Co2 sites are not the same.\cite{EPL2009} We have
suggested for long that the Co1 sites had an isotropic NMR orbital shift
which is a strong signature that the Co1 sites are themselves nearly Co$%
^{3+} $, the orbital shift magnitude being furthermore nearly identical to
that found in the band insulator phase Na$_{1}$CoO$_{2}$,\cite%
{LangNa1,MHJulienNa1} in which all Co sites are non-magnetic Co$^{3+}$. More
accurately, as we are dealing with a metallic system the filling of the
bands involving the Co $t_{2g}$ orbitals at the Fermi level is quite
different on the Co1 and Co2 sites, and we can expect that the hole filling $%
\epsilon $ of the $t_{2g}$ orbitals is small on the Co1 site. Let us assume
that the hole content on Co2 sites is $\delta $. As the number of holes per
CoO$_{2}$ formula unit is 1/3, the charge neutrality for the kagom\'{e}
structure implies then $(3\epsilon +9\delta )/12=1/3$, that is
\[
\delta =4/9-\epsilon /3.
\]

\subsubsection{(a) Isotropic spin susceptibilities and $T_{1}$ data}

The $T$ dependent NMR shifts of the Co sites, which scale with that of $%
^{23}$Na give a measure of the spin susceptibilities on the Co sites. If we
consider in Table~\ref{tab:CoShifts} the isotropic shifts $K_{iso}=(1/3)\sum
K_{\alpha \alpha }$ on the Co sites, we immediately find that the average
values for the two Co1 sites $K_{iso}($Co1$)/^{23}K\approx 1.3$ is an
order of magnitude smaller than that for the Co2 sites $K_{iso}($Co2$%
)/^{23}K\approx 10.6$, which indicates that the unpaired spins are
dominantly on the Co2 sites. Similarly, both from NMR and low $T$ NQR data we did find a ratio $\approx $70 between the spin contributions to $%
1/T_{1}$ for the two sites. Assuming similar hyperfine couplings for the
two sites, one could anticipate from these results a ratio of $\sqrt{70} \approx 8$ of the local spin susceptibilities on the two sites, that is $\epsilon /\delta \lessapprox 1/8$. \textit{This corresponds to an upper limit of 0.06 for epsilon.}

\subsubsection{(b) Isotropic orbital susceptibilities and orbital
contribution to $(T_{1}T)^{-1}$}

Let us consider the difference found for the orbital shifts between the Co1
and Co2 sites. While for Co1 sites $K^{orb}$ was found isotropic already in
Ref.~\onlinecite{CoPaper}, we do confirm here that $K^{orb}$ is anisotropic
in plane for the Co2 sites with a sizable increase of $K_{iso}^{orb}$. We
can therefore consider that the contribution of the hole orbitals to the
orbital susceptibility is responsible for the deviation of $K_{iso}^{orb}$
with respect to 1.91(3)\% found in Na$_{1}$CoO$_{2}$.\cite{LangNa1} In a
rough order of magnitude comparison we might consider then that the hole
density on the Co sites is proportional to $K_{iso}^{orb}-K_{iso}^{orb}(x=1)$, as we already suggested in Ref.~\onlinecite{CoPaper}. We do find then,
from the results of Table~\ref{tab:CoShifts} that $K_{iso}^{orb}($Co$%
1)-K_{iso}^{orb}(x=1)=-0.01\pm 0.05(\%)$, while $K_{iso}^{orb}($Co$%
2)-K_{iso}^{orb}(x=1)=0.54(\%)$. This would give again an upper estimate of $\epsilon /\delta \lesssim 1/10$. Similarly knowing \cite{LangNa1} that for
Co$^{3+}$ in insulating Na$_{1}$CoO$_{2}$ the spin lattice relaxation rate
is very weak, one could anticipate that the metallic orbital contributions
to $(TT_{1})^{-1}$ would be linked to the hole doping on the corresponding
site. So there is no surprise to find out that$\ (TT_{1}^{orb})^{-1}$ is
much larger for Co2 than for Co1.

So the data lead us to consider that we may neglect $\epsilon $ as a first
approximation and keep $\delta \approx 0.44$, although only accurate band
calculations could tell exactly whether the partly filled bands at the Fermi
level do not include any partial Co1 hole orbital in the kagom\'{e}
structure.

\subsubsection{(c) EFG induced by charge order}

Let us consider now the quadrupole frequencies $\nu _{Q}$, which are
directly proportional to the EFG on the nuclear site probe. The latter
arises from a non-symmetric distribution of electric charges around it.
These charges can originate from non-bonding electrons, electrons in the
bonds and charges of neighboring atoms or ions. The components of the EFG
tensor can be written as the sum of two terms, the lattice $V_{\alpha \alpha
}^{\mathrm{latt}}$ and electron $V_{\alpha \alpha }^{\mathrm{el}}$
contributions:
\begin{equation}
V_{\alpha \alpha }=(1-\gamma _{\infty })V_{\alpha \alpha }^{\mathrm{latt}%
}+(1-R_{el})V_{\alpha \alpha }^{\mathrm{el}},  \label{eq:Vaa}
\end{equation}%
where parameters $\gamma _{\infty }$ and $R_{el}$ are the Sternheimer
antishielding factors, which characterize the enhancement by the core
electrons of the atom of the EFG on the bare nucleus with respect to the EFG
due to the outer electron distributions.

The first contribution arises from all ion charges outside the ion under
consideration and can be calculated in an approximation assuming point
charges
\begin{equation}
V_{\alpha \alpha }^{\mathrm{latt}}=\sum_{i}\frac{q_{i}(3cos^{2}\theta _{i}-1)%
}{r_{i}^{3}},  \label{eq:Vbb}
\end{equation}%
where $q_{i}$ is the charge of ion $i$, located at distance $r_{i}$ of the
probe nucleus, in a direction at an angle $\theta _{i}$ from the Z axis of
the EFG tensor.

The second term in Eq.~\ref{eq:Vaa} arises from unfilled electron shells of
the orbitals of the considered site and of the distortions of the inner
electron orbitals that they induce, embedded in the known value of ($%
1-R_{el})$.

We have then to understand the relative weight of these two contributions in
the measured EFG in this particular case of cobaltates. For a fully filled $%
t_{2g}$ multiplet such as the Co$^{3+}$ low spin state, the ionic structure
is isotropic and the second term in Eq.~\ref{eq:Vaa} should contribute
negligibly to the on site EFG.

Let us then point out that we can deduce significant indications about the
respective weights of these two terms by considering the point charge
calculations done in Ref.~\onlinecite{H67NQRprb}. There we deduced average values for the Co1a and Co1b sites of $\nu _{Q}=1.32$~MHz and a similar value $\nu _{Q}\approx 1.48$~MHz for the average of the Co2a and Co2b sites.
Experimentally the corresponding respective values $\nu _{Q}\approx $
1.29~MHz and $\nu _{Q}\approx $2.36~MHz differ significantly. This leads us
to suggest then that the excess contribution of about 0.9~MHz on the Co2
sites is due to the on site hole orbitals. Furthermore, although we cannot
rely fully on the numerical values of the point charge calculation, the
agreement found with the Co1 data is indeed quite compatible with a
negligible contribution of the on site orbitals on the Co1 sites, that is
with a very small $\epsilon $ value.

To check the consistency of this analysis we may compare the data with that
obtained for $x=0.35$, for which we found that the Co sites are uniformly
charged.\cite{CoPaper} In that case, assuming a simple ordered distribution
of Na2 sites and $x=1/3$, the point charge calculations give values of $\nu
_{Q}\approx $ 2.5~MHz for all Co sites, whatever their location with respect
to the Na sites, while the NMR data of Ref.~\onlinecite{CoPaper} corresponds to $\nu_{Q}\simeq $ 4.1~MHz. So this much larger value would indicate that a
1.6~MHz on site contribution should be associated to $2/3$ hole per Co site.
Such a contribution to $\nu _{Q}$ of 2.4~MHz/hole would correspond to
1.1~MHz for 0.44 holes on the Co2 sites in the $x=2/3$ phase, remarkably
close to the 0.9~MHz value estimated above.

So, although these point charge estimates are certainly not fully reliable,
they indicate that \textit{the excess EFG found on the Co2 sites has the right order of magnitude and is compatible with the poor hole occupancy of the Co1
sites}, as obtained here-above from the analyzes of $K^{iso}$ and $K_{orb}^{iso}$.

\subsection{In plane anisotropies and orbital order}

\begin{figure}[tbp]
\center
\includegraphics[width=1\linewidth]{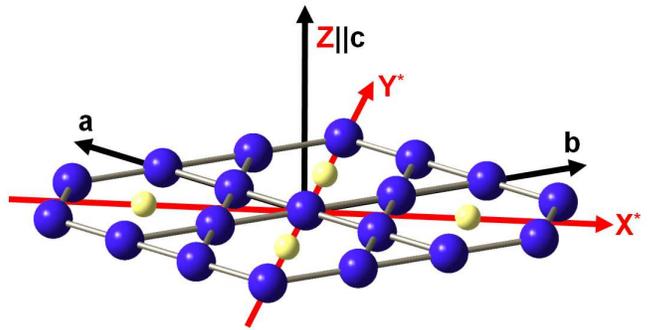}
\caption{(Color online) The kagom\'{e} structure in the CoO$_2$ planes
allows to clearly locate two orthogonal axes ($X^{\ast }$, $Y^{\ast }$)
which should be the Co2 site principal axes in the plane.}
\label{fig:NewFig}
\end{figure}

The simple consideration of the atomic kagom\'{e} structure displayed in
Fig.~\ref{fig:NewFig} allows one to expect an asymmetry of the local
properties at the Co2 site, and to locate the two orthogonal axes ($X^{\ast
},Y^{\ast }$) which should be the Co2 site in plane local principal axes.

Concerning the electronic structure, it is quite well known that the
metallic bands of cobaltates are built from the Co $t_{2g}$ orbitals which
subdivide in the $a_{1g}$ orbital which is axial and perpendicular to the
plane and two subsets of in plane $e_{g}^{\prime }$ orbitals. From LDA
calculations of the band structure, done so far for uniformly charged
cobaltate planes (see for instance Ref.~\onlinecite{PhysRevB.78.012501}), one usually expected a Fermi surface containing an $a_{1g}$
sheet at the Fermi level centered at the $\Gamma $ point and $e_{g}^{\prime
} $ pockets at the zone corners.\cite{SinghPRB61} Only the larger $a_{1g}$
electron pocket has been observed so far by ARPES experiments, and that
whatever the hole content.\cite{YangPRL95,VBrouetEPL} It has often been
considered that, even in such simple uniform representations of the CoO$_{2}$
plane, the electronic correlations could largely influence the electronic
structure by narrowing the bands and possibly pushing the $e_{g}^{\prime }$
hole orbitals below the Fermi level.\cite{LiebschIshidaEPJB}

All that might not be relevant to the Na ordered and charge
disproportionated structures, which would give a smaller Brillouin zone and
a larger number of possible bands at the Fermi level. But surprisingly
here we find no experimental sign for multiband behaviour. In any
case the dominant magnetic Curie-Weiss behaviour is governed by a
single band in which correlations are important. So far the nature of the
Co orbitals which participate in this correlated electronic band are not
readily available from the data. It remains then important to try to
determine experimentally what are the respective weights of the $a_{1g}$ and
$e_{g}^{\prime }$ orbitals involved at the Fermi level in the actual
electronic structure of the 2/3 phase.

We consider that the non axial NMR parameters determined here may help to
answer this question. Assuming that Co1 is in a completely filled shell Co$%
^{3+}$ ionic state (that is $\epsilon $=0), if the electronic structure only
results from $a_{1g}$ hole orbitals on the Co2 sites, the non axial NMR
parameters would only be induced by the absence of transfer integral between
Co2 and Co1 sites. But if a fraction of the hole density on Co2 resides on $%
e_{g}^{\prime }$ orbitals, the local asymmetry of the transfer integrals
would induce distinct occupancies of the two $e_{g}^{\prime }$ orbitals on
the Co2 sites. One could expect a larger hole occupancy of the $e_{g}^{\prime }$ orbital pointing in the $X^{\ast }$ direction in Fig.~\ref%
{fig:NewFig}, which would correspond to an ordering of the in plane hole
orbitals on the Co2 sites. In any case such an unbalanced population of the $%
e_{g}^{\prime }$ orbitals would induce an on site anisotropy of the NMR
parameters on Co2 sites. We shall here try to estimate the $a_{1g}$ and $%
e_{g}^{\prime }$ hole populations by considering first the asymmetry of the
EFG and then that of the NMR shifts.

\subsubsection{In plane EFG anisotropy on Co2 sites}

Both asymmetries of the distributions of Na ions and Co charges located in
the plane do introduce an in plane EFG asymmetry on the Co2 sites. We could
evaluate the part of this asymmetry due to ionic charges from the point
charge calculations of the first term of Eq.~\ref{eq:Vaa} done in Ref.~%
\onlinecite{H67NQRprb}. We indeed found that the distribution of Na$^{+}$
charges induces a slight tilt of the $Z$ principal axis of the EFG with
respect to the $c$ axis on both Co2 sites. The overall ionic charge
distribution also results in $X$ and $Y$ principal axes of the EFG which are
distinct on the Co2a and Co2b sites and which were found markedly modified
if we displace the Na atoms with respect to their ideal positions as
suggested by the x-ray Rietveldt data analysis of Ref.~\onlinecite{H67NQRprb}. However in all cases considered, the computed values of $\left\vert
V_{YY}-V_{XX}\right\vert $ was found to correspond to a difference of NQR
frequencies $(\nu _{Y}-\nu _{X})$ in a range between 0.4 and 0.9~MHz. This
appears on the average significantly smaller than the 0.8 and 0.9~MHz values
deduced from the experimental asymmetry parameter $\eta \approx 0.35$ found
for both Co2a and Co2b sites (see Table~\ref{tab:CoSimParam}).

We may then suggest that some of the non axial charge distribution is linked
with the on site term. As the $a_{1g}$ orbital is axial, any on site non
axial contribution to the EFG of the Co2 could only result from an unbalance
in the populations of the $e_{g}^{\prime }$ orbitals.\cite%
{MBLepetitPrivateComm, JSoret}. The fact that the asymmetry parameter and
EFG on the two Co2 sites are quite similar further supports the existence of
a sizable on site $e_{g}^{\prime }$ hole orbital contribution to the EFG.
Unfortunately the large sensitivity of the point charge calculations of the
EFG to the actual atomic structure does not allow us to subtract reliably
this EFG contribution from the experimental data, and to deduce then the
principal axes and magnitude of the on site contribution to the EFG. It is
clear then that more accurate EFG determinations based on ab initio band
calculations are then needed to relate the EFG data to the relative
populations of the $a_{1g}$ and $e_{g}^{\prime }$ orbitals.

\subsubsection{In plane Knight shift  anisotropies}

While the anisotropy of Na charge distribution contributes to the EFG
anisotropy, the Na ions are not involved directly in the magnetic
properties, which are fully associated with the CoO$_{2}$ layers. So, the
anisotropies of the NMR shifts of the Co2 sites should give much better
indications on the anisotropy of the in plane electronic structure. The non
axial values of the Co2 shifts summarized in Table~\ref{tab:CoShifts} are
indeed pointing out the existence of an in plane electronic anisotropy. One
can notice there that both $K^{orb}$ and $K^{spin}(T)$ are much larger in
the $X$ direction than in the $Y$ direction. Let us note that these
experimental directions are those identified from the EFG\ tensor, with the
usual convention $\left\vert V_{YY}\right\vert >\left\vert V_{XX}\right\vert
$, so that the smallest quadrupole frequency occurs in the planar direction
for which the NMR shifts are largest, as seen directly in the spectra.
However so far we do not know for sure how $(X,Y)$ relate to the $(X^{\ast
},Y^{\ast })$ axes defined in Fig.~\ref{fig:NewFig}.

We did find as well an anisotropy of the spin contributions to $(TT_{1})^{-1}$ which correlates with the anisotropy of hyperfine couplings, so that no
really new information is embedded in this result, especially on the in
plane anisotropy.

\paragraph{Orbital shift in plane anisotropy}

SQUID data had allowed us to evidence that the macroscopic susceptibility is
larger along the $ab$ plane than in the $c$ direction. This data allowed us
to establish that the $T$ dependence of the susceptibility, which is
determined by the spin contribution $\chi ^{\mathrm{spin}}$ $(T)$ is nearly
isotropic, which has been confirmed as well from $^{23}$Na NMR data.\cite%
{NaPaper} So the susceptibility anisotropy is dominated by that of $\chi
^{orb}$ and $\chi _{ab}^{orb}>\chi _{c}^{orb}$ is sufficiently large to be
practical to align the single crystal grains in the applied field at room
temperature. But these SQUID data on powder crystallite samples do not give
any indication about the in plane anisotropy of orbital susceptibility. The
present NMR data in Table~\ref{tab:CoShifts} establishes a planar anisotropy
of $\chi ^{orb}$ on the Co2 sites, for which $K_{X}^{orb}>K_{Y}^{orb}>K_{Z}^{orb}$.

Let us point out that $K^{orb}$ can be a purely ionic term, as is the case
for Co$^{3+}$, but states at the Fermi level might also give a specific
contribution. This is indeed expected as we found such a metallic band
orbital contribution to $(T_{1}T)^{-1}$ for both Co1 and Co2 sites. In fact $%
K_{Z}^{orb}$ is unmodified with respect to that on Co$^{3+}$, so that the
extra contributions found for $K_{X}^{orb}$ and $K_{Y}^{orb}$ could be
mainly due to such band contributions, as is suggested by the correlation
between the spin and orbital in plane anisotropies found in Table~\ref%
{tab:CoShifts}. One could notice however hereagain that if the hole orbitals
were purely $a_{1g}$, one would not expect any in plane anisotropy of $%
K^{orb}$, so that heregain band calculations of $\chi _{\alpha \alpha
}^{orb} $ might help to determine the populations of the $a_{1g}$ and $%
e_{g}^{\prime }$ orbitals.

These observations on the Na and Co sites are reminiscent of the respective
situations found for the NMR of the $^{89}$Y and of the $^{63}$Cu nuclei of
the CuO$_{2}$ planes of the YBCO cuprates. In that case, as for Na NMR here,
the $^{89}$Y NMR shift which is only coupled to spin magnetism was found
nearly isotropic\cite{AlloulYBCO} while the anisotropy of orbital shift of $%
^{63}$Cu was due to that of $\chi ^{orb}$, though in that case the orbital
contribution due to the $d_{x^{2}-y^{2}}$ in plane hole orbital\cite%
{TakigawaYBCO} does not display any anisotropy in the CuO$_{2}$ plane.

\paragraph{Spin shift in plane anisotropy}

As the spin susceptibility is nearly isotropic from SQUID and Na NMR shift
data, the large in plane anisotropy of the Co2 NMR spin shift data $%
K_{\alpha }^{spin}(T)$ has to be associated with an anisotropy of hyperfine
couplings, which are linked with the occupancy of the electronic orbitals involved at the Fermi level.

Further comparison with the cuprates could help us to better understand the
in plane anisotropy of $^{59}$Co NMR spin shifts detected here. Quite
generally the spin shift tensor $K_{\alpha \alpha }^{\mathrm{spin}}$ is
slightly more complicated than indicated in Eq.~\ref{eq:Kcomp} as nuclear
spins at site $i$ can be coupled both to the on site magnetism and to that
of their $j$ near neighbour sites, so that
\begin{equation}
K_{\alpha \alpha }^{\mathrm{spin}}(i)=A_{\alpha \alpha }\frac{\chi _{\alpha
\alpha }^{\mathrm{spin}}(i)}{g\mu _{B}}+\sum_{j}B_{\alpha \alpha }\frac{\chi
_{\alpha \alpha }^{\mathrm{spin}}(j)}{g\mu _{B}},
\end{equation}%
where $A_{\alpha \alpha }$ is the anisotropic on-site hyperfine coupling
tensor and $B_{\alpha \alpha }$ is the transferred hyperfine coupling
through oxygen orbitals with the $j$ nearest neighbors. In YBCO there is no
hole occupancy on the Y site and therefore no on site hyperfine coupling $%
A_{\alpha \alpha }$ so that the hyperfine coupling of $^{89}$Y to the Cu
site magnetism is only due to the transferred $B_{\alpha \alpha }$ which has
been found isotropic.\cite{AlloulYBCO} On the contrary the anisotropy of on
site coupling $A_{\alpha \alpha }$ is responsible for the large measured $%
(c,ab)$ anisotropy of the $^{63}$Cu spin shift. Even in this simple case
with a single Cu axial site it took sometime to understand that a rather
large isotropic value of $B_{\alpha \alpha }$ was required to explain the
data.\cite{MilaRice} Here, we shall benefit of this approach to perform a
tentative preliminary analysis of the Co1 and Co2 data summarized in Table~%
\ref{tab:CoShifts}.

For the axial Co1 sites, only a $(c,ab)$ shift anisotropy is detected. If $\epsilon $=0, then Co1 sites are non-magnetic Co$^{3+}$ ions which bear no on
site magnetic spin susceptibility. As the Y ions in the cuprates they would
be inert magnetically and would sense the magnetism of the Co2 orbitals only
through transferred hyperfine couplings $B_{\alpha \alpha }$ with the Co2
magnetism. The $(c,ab)$ anisotropy should then be associated with an
anisotropy of the coupling $B_{\alpha \alpha }$ of the Co1 with its first
Co2 nearest neighbours. A dipolar coupling of Co1 with its six Co2
neighbours would correspond to a larger negative contribution to the shift
for $H\parallel c$, contrary to observation. A non zero $\epsilon $ value
with a small filling of the $a_{1g}$ hole orbital on the Co1 would better
explain the sign of the $(c,ab)$ anisotropy with a maximum shift
contribution for $H\parallel c$.

But of course the most intriguing experimental information resides in the
totally non-axial spin shift values found here for the Co2 sites, given in
Table~\ref{tab:CoShifts}. Averaging the data for the Co2a and Co2b sites
results in $K_{X}^{spin}/^{23}K=18.9$; $K_{Y}^{spin}/^{23}K=7.5$ and $%
K_{Z}^{spin}/\ ^{23}K=4.9$.

These large anisotropic contributions to the shift might be dominated by on
site hyperfine couplings with the $a_{1g}$ and $e_{g}^{\prime }$ orbitals.
We can note hereagin, as for the discussion of the EFG, that the axial
symmetry of the $a_{1g}$ which points along the $c$ axis would only yield
axial shift contributions and would not differentiate $X$ and $Y$
directions, contrary to the $e_{g}^{\prime }$ orbitals. So the large
difference between $K_{X}$ and $K_{Y}$ can only be attributed to the in
plane order of $e_{g}^{\prime }$ orbitals associated with the kagom\'{e}
structure. If one considers the transfer integral paths of $e_{g}^{\prime }$
orbitals between Co2 neighbouring sites, one would be inclined to associate
the $X$ direction, with the largest shift value with the direction $X^{\ast
} $ sketched in Fig.~\ref{fig:NewFig}.

Let us note however that nothing forbids as well to consider here that the $%
B_{\alpha \alpha }$ term itself is anisotropic. Detailed
consideration of the Wannier orbitals constructed for the full atomic
structure of this ordered $x=1/3$ phase could help to explain these measured
anisotropies and would allow altogether to determine the relative
populations of $a_{1g}$ and $e_{g}^{\prime }$ hole orbitals on the Co2 sites
at the Fermi level.

\section{Conclusion}

In this experiment we have completely determined experimentally the NMR
parameters of the four Co sites of the atomic structure of this Na$_{2/3}$CoO%
$_{2}$ phase, which allows us to confirm that its electronic structure
differentiates in fact two sites in the CoO$_{2}$ plane: Co1 with a charge $%
3+\epsilon $ and Co2 with a charge $3.44-\epsilon $. The data for the EFG,
the orbital NMR shifts and orbital $T_{1}T$ are compatible with a hole
content at most of $\epsilon \lesssim 0.06$ on the Co1 site as we had
already anticipated.

From our NMR data we confirm as well that a single band dominates the
anomalous electronic properties of this phase at the Fermi level. A totally
unexpected aspect revealed by these data is the existence of large in plane
anisotropies of the EFG and NMR shifts on the Co2 site. We have shown that
this implies that the hole orbitals on those sites are not exclusively on $%
a_{1g}$ orbitals and that parts of the holes reside in the $e\prime _{g}$
in plane orbitals which should then play an important role in the
delocalization of the holes.

So the single hole band hybridizes those orbitals and the kagom\'{e} like
electronic structure that we evidenced then displays not only a charge
disproportionation between Co1 and Co2 sites, but as well an in plane
charge order of the $t_{2g}$ orbitals on the Co2 sites. Quantum chemistry
calculations taking into account the full structure might permit to go
beyond the qualitative arguments developed here and should provide
quantitative data for the $t_{2g}$ energy levels from the values obtained
experimentally for the EFG, hyperfine couplings and orbital shift parameters.

One would like then to understand the respective roles of the Na order and of
the electronic correlations in driving this electronic structure. In
principle LDA calculations including the Na order should allow to understand
whether the Na potential is sufficient to shift down the Co1 $t_{2g}$ energy
levels and to induce the Co$^{3+}$ filling of the Co1 orbitals. Preliminary
calculations along this line seem to indicate that this is not the case and therefore that it is essential to better take into account the electronic interactions.\cite{FLechermannPrivateComm} In such a case the kagom\'{e} like
organization of the electronic structure would be an intrinsic
disproportionation in the Co planes which would lock the Na order.\cite{BoehnkeLechermann}

Our work evidences that the resulting situation is quite simple in this Na$%
_{2/3}$CoO$_{2}$ phase, and resumes in a simple two site charge
disproportionated CoO$_{2}$ plane which could be introduced in a simplified
model. This would allow one to perform then cluster DMFT calculations to
establish the role of correlations which are of course essential to explain
the large Curie-Weiss like $T$ dependence found for the susceptibility as
well as the 2D ferromagnetic correlations deduced from our spin lattice
relaxation data.\cite{EPL2008}

In any case we have shown that the Na$_{x}$CoO$_{2}$ compounds display
stable Na ordered phases  in which charge ordered states of the CoO$_{2}$
planes occur so that the real atomic structure has much larger 2D unit
cells than that of the CoO$_{2}$ plane. We might anticipate then that most
ARPES experiments, which isolate a specific surface after cleaving do not
sample the electronic properties of the actual 3D ordered state but are
affected by the unavoidable Na disorder in such experimental conditions,
although in some cases specific local Na orderings have been observed by
STM experiments on cleaved surfaces.\cite{PhysRevLett.100.206404}

An interesting case is that of some misfit cobaltates \cite{MaignanMisfits} for which the multilayer structure is obtained by stacking rocksalt structures with CoO$_{2}$ planes. In that case cleaving is done in the middle of the rocksalt
structure, as in High $T_{c}$ cuprates, and an ARPES experiment \cite{VBrouetEPL} has evidenced that in the specific case of BaBiCoO$_{2}$ the electronic structure displays a reconstruction which could result from a charge order analogous to that seen in the Na$_{x}$CoO$_{2}$ phases. Unfortunately in
that case it has so far not been possible to perform NMR experiments
allowing to resolve the sites,\cite{BobroffMisfits} so that detailed comparisons are not possible.

Finally we have established here on a specific case the importance of charge
disproportionation and orbital ordering in this compound. No doubt that for
all hole dopings for $x>0.5$ similar effects do occur as diverse Na
orderings are anticipated and observed and result in very distinct
correlated electronic states.\cite{Mendels05,Shu,EPL2008,LangNFD} We think that the present NMR studies allow us to reach for the first time a very detailed description of such effects in multiorbital transition oxides.

\begin{acknowledgments}
We would like to thank here A.V.~Dooglav and T.A.~Platova for the
stimulations provided by the NQR experiments, J.~Bobroff and P.~Mendels
for their help on the experimental NMR techniques and for constant
interest and stimulating discussions, and A.N.~Lavrov for helpful discussions on material properties aspects. We acknowledge as well G.~Collin for his initial unvaluable help on sample characterization and x-ray analysis.
We benefited from enlightening discussions with M.B.~Lepetit and J.~Soret
about the quantum chemistry of these systems and with F.~Lechermann about
the EFG and the band structure of the cobaltates.

I.R.M. thanks for partial support of this work the Russian Foundation for Basic Research (project no.10-02-01005a), Ministry of Education and Science of the Russian Federation (project no.2010-218-01-192 and theme no.1.44.11) and Universit\'e Paris-Sud for associate professor visiting positions. This work has been done within the Triangle de la Physique and H.A. thanks its initial funding by ANR grant Oxyfonda NT05-4 41913.
\end{acknowledgments}

\appendix

{\color{red}

\section{Oxygen vacancies}

\label{TaiwaneseAreWrong}

A recent publication \cite{TaiwanPRB2010} is dedicated to discredit our
results \cite{EPL2009,H67NQRprb} and in support of structural models of the Na
cobaltates that this group has been proposing for many years for large $x$
values.

(i) In this paper the authors suggest that our samples, being synthesized at
high $T$, are not Na$_{2/3}$CoO$_{2}$ but rather Na$_{2/3}$CoO$_{1.98}$. We
can hardly discard solely from our NMR data the presence of such a low
content of oxygen vacancies. As each Co has six near neighbor oxygen, only
about 6\% of the Co would have such near neighbor vacancies. If those sites
would get well defined NMR shift and EFG they should be detected either in
NMR or NQR. But as soon as some disorder is expected in the corresponding
parameters due to long distance interactions between defects, we would not
be able to detect them. We may however easily exclude that vacancies could
induce perturbations beyond their first nearest neighbors. Those would yield
a broad background signal involving then a large Co NMR intensity. So, if
present, such oxygen vacancies should just play a similar role to that of Na
atoms and give delocalized carriers on the kagom\'{e} structure,
independently of the Co1 sites Co$^{3+}$ signal which are connected with the
ordered Na sites.

To conclude on that point, there is so far no \textit{microscopic}
experimental evidence for the existence of such oxygen vacancies neither in
Ref.~\onlinecite{TaiwanPRB2010} nor in our NMR/NQR data.

(ii) Beyond this materials research aspect, Shu \textit{et al.} speculate, heragain without any \textit{microscopic} evidence, that the Na order in our samples (putatively Na$_{2/3}$CoO$_{1.98}$) might be a structure built from a
succession of planes with Na trivacancies and Na tetravacancies,\cite%
{TaiwanPRB2009} which they have proposed for their sample with an assumed
composition of Na$_{0.71}$CoO$_{2}$.

We claim that this is  definitely contradicted by all our former NMR, NQR
and x-ray data \cite{EPL2009,H67NQRprb} and by the present $^{59}$Co NMR
data. It is easy to establish that this structure corresponds to $5$ Na
crystallographic sites (with relative intensities 3/6/3/2/3), and $8$ Co
sites (with relative intensities 1/1/1/3/3/3/3/9). This is well beyond the three Na sites we have detected by NMR \cite{NaPaper} and the four Co sites we have detected both by NQR \cite{EPL2009,H67NQRprb} and here by NMR. Finally
the structure we have proposed from the analysis of our NMR/NQR data in
Ref.~\onlinecite{EPL2009} has been confirmed by the perfect fit obtained with the Rietveldt x-ray data analysis performed on our samples.\cite{H67NQRprb}

So the self consistency of our data on our Na$_{2/3}$CoO$_{2}$ samples
cannot be put into question. Whether the structure proposed by Shu \textit{et al.} for their Na$_{0.71}$CoO$_{2}$ samples is the actual one remains a question which might be resolved in the future by performing independent NMR/NQR investigations on their samples.
}

\section{Check of the powder alignment by NMR}

\label{AppendixPowdAlign}

The alignment of the sodium cobaltates powders being dependent on the
microstructure of the grains, some of our samples were indeed not well
oriented. We demonstrate here how we could use the NMR spectra to
distinguish poorly oriented samples from well oriented ones.

\begin{figure}[tbp]
\center
\includegraphics[width=0.8\linewidth]{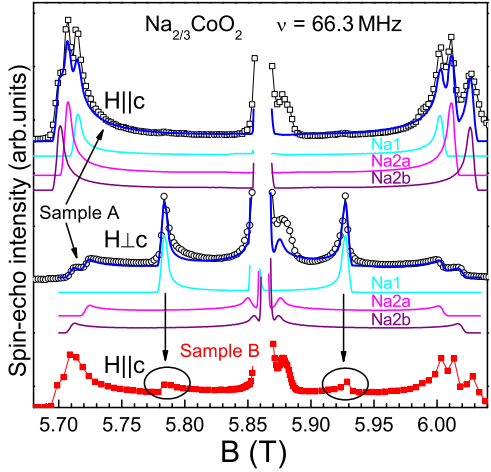}
\caption{(Color online) $^{23}$Na NMR spectra: comparison of oriented sample
A (black open circles for $H_0 \perp c$ and open squares for $H_0 \parallel
c $) and sample B (red closed squares for $H_0 \parallel c$). The solid
colour lines display the computed spectra for the three different Na sites,
and the total spectum which fits perfectly the data for sample A. The
presence of the Na1 singularities specific for the $H_0 \perp c$ direction
spectrum of sample A in the $H_0 \parallel c$ spectrum of sample B is clear
and allowed us to conclude that sample A is better oriented than sample B. The extra line near the central transition at about 5.88~T with small Knight shift has a very weak intensity, which corresponds at most to 5\% of all Na nuclei, and might be assigned to defect sites in the ordered Na structure, or to a minute amount of impurity phases present in these samples with large crystallites.}
\label{fig:NaBad}
\end{figure}

In Fig.~\ref{fig:NaBad} we show a $^{23}$Na NMR spectra measured in two
samples of Na$_{2/3}$CoO$_{2}$ compound. One of them is well oriented sample
- Sample A - and we show there the spectra measured in two directions of the
applied magnetic field - $H_{0}\perp c$ and $H_{0}\parallel c$. As evidenced
before \cite{NaPaper,H67NQRprb} in the Na$_{2/3}$CoO$_{2}$ phase there are
three unequvalent sodium sites - Na1, Na2a, Na2b. We found that although the
central lines and satellites singularities of the $^{23}$Na spectrum can be
easily located, the overall shape of the NMR spectrum is very sensitive to
the degree of orientation of the powder sample. In Fig.~\ref{fig:NaBad} we
also show results of the computer simulations of the $^{23}$Na NMR spectra
using the parameters of Eq.~\ref{eq:Hamilt2} from Ref.~%
\onlinecite{NaPaper}. Here we have optimized the particles $c$ axis angular
orientation distribution to fit accurately the spectrum (Lorentzian
distribution of $\pm 5^{\circ }$ width). It's important to mention that we
used exactly the same angular distribution parameters for the $^{59}$Co
simulations shown in Figs.~\ref{fig:GrCo1Sim5K}-\ref{fig:GrCo2Sim60K}. As
one can see the agreement between experimental and simulated spectra are
very good.

At the bottom of Fig.~\ref{fig:NaBad} we show a $^{23}$Na NMR spectrum
measured in sample B of Na$_{2/3}$CoO$_{2},$ which happened to be less well
oriented. Comparing this spectrum with that measured in Sample A it is easy
to see for $H_{0}\parallel c$, the presence of Na1 singularities specific to
the $H_{0}\perp c$ direction of applied magnetic field. That is due to a
fraction of particles which remained randomly oriented as they were either
not single crystallites, or were too much packed or agglomerated and could
not orient. Also a less prominent, but still detectable feature of this less
oriented sample is the poor resolution of the quadrupolar satellites.

\begin{figure}[tbp]
\center
\includegraphics[width=1\linewidth]{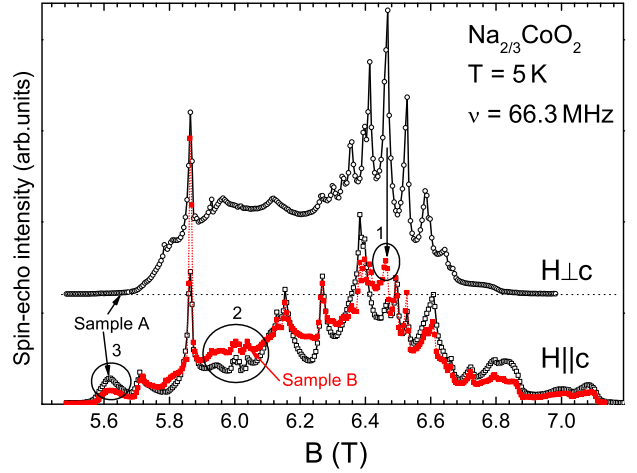}
\caption{(Color online) Comparison of $^{59}$Co NMR spectra of the well
oriented sample A (black open circles for $H_{0}\perp c$ and open squares
for $H_{0}\parallel c$) with that of the less well oriented sample B (red
closed squares for $H_{0}\parallel c$). The extra contributions in the $%
H_{0}\parallel c$ spectrum of sample B due to the Co1 and Co2 signals of
non-oriented particles are highlighted within circles (1) and (2).}
\label{fig:CoBad}
\end{figure}

In Fig.~\ref{fig:CoBad} we demonstrate the distinction between the $^{59}$Co
NMR spectra measured in these two samples as well. The comparison can
summarized as follows

1) First of all in the sample B spectrum measured for $H_{0}\parallel c$
one can see an extra line, circled as (1) which corresponds to the central lines of Co1a and Co1b sites in the $H_{0}\perp c$ direction - marked by arrow in Fig.~\ref{fig:CoBad}.

2) Also one can see that for the well oriented sample A that the NMR spectrum
measured for $H_{0}\parallel c$ is rather symmetric whereas the spectrum for
$H_{0}\perp c$ is asymmetric with a higher signal intensity for the lower
field values which could be assigned to fast relaxing cobalts. In sample B
even in the $H_{0}\parallel c$ spectrum some asymmetry could be seen and is
here again associated with the contribution from grains which did not orient
in the field. This extra intensity, circled as (2), had mislead us by
suggesting that a Co site labeled as Co3 could have a large purely axial
shift (see Fig.~2c in Ref.~\cite{CoPaper}).

3) Again in the sample B spectrum the quadrupolar satellites are less
resolved as one can see in Fig.~\ref{fig:CoBad}, on the part circled as (3).

\section{Determination of $K(T)$ from full analyzes of the $^{59}$Co NMR
spectra}

\label{AppendixCo2XY}

The $^{59}$Co NMR of the slow relaxing Co1 sites in the Na$_{\approx 0.7}$CoO%
$_{2}$ compounds are easy to observe and were seen even by continuous wave
NMR.\cite{Ray} In the Na$_{2/3}$CoO$_{2}$ compound even small details for
the Co1a and Co1b sites could be detected and have been reported.\cite%
{GavilanoCo07} The axial symmetry of these sites symmetrizes the magnetic
shift and quadrupolar tensors which makes the NMR lines of $^{59}$Co in
these positions rather narrow and intense.

The case is more complicated for the fast relaxing Co2 sites in the Na$%
_{\approx 0.7}$CoO$_{2}$ compounds. Even in the single crystals the NMR
spectra of these sites are broad and not well resolved.\cite{ImaiPRL1} The
same situation occurs in our oriented powder samples - see Fig.~\ref%
{fig:GrCoExp} as an example. As in our samples the $c$-axis of different
crystallites are aligned along $H_{0}\parallel c$ one can distinguish in the
spectrum the two sets of 7 lines which correspond to the different
transitions $m\leftrightarrow (m-1)$ for the two cobalt sites Co2a and Co2b.
For $H_{0}\perp c$, the NMR spectrum of the fast relaxing cobalts is found
broad and non-resolved, as the particles in our samples have a powder
distribution of $ab$ axes for $H_{0}\perp c$. This experimental fact
immediately tells that the local symmetry for the Co2 sites is lower than
for the Co1 sites.

To clarify the spectra of the Co2 sites for $H_{0}\perp c$ we used various
facts:

1) The position of the different singularities in the fast relaxing Co2
spectra are the same whatever the sample and whatever the sample preparation
method, as one can see in the Fig.~\ref{fig:GrCoExp}. The positions of the
singularities in the NMR spectra depend on both the quadrupole splitting
(which could be considered practically as temperature independent) and the
magnetic shift which scales with the susceptibility $\chi _{s}(T)$ .

2) The main feature of the Na$_{2/3}$CoO$_{2}$ phase is a large temperature
dependence of the spin susceptibility. As we have shown that the local
fields are quite similar for the three detected Na sites \cite{NaPaper}, the
first moment (center of gravity) of the $^{23}$Na NMR spectrum $%
^{23}K_{s}^{iso}=A\chi _{s}(T)$ allows us to follow the spin susceptibility $%
\chi _{s}(T)$ of the cobalts up to room $T$. Therefore using $%
^{23}K_{s}^{iso}(T)$ as a reference we tried to follow the corresponding $T$
variation of the Co shift tensor.

\begin{figure}[tbp]
\center
\includegraphics[width=1.0\linewidth]{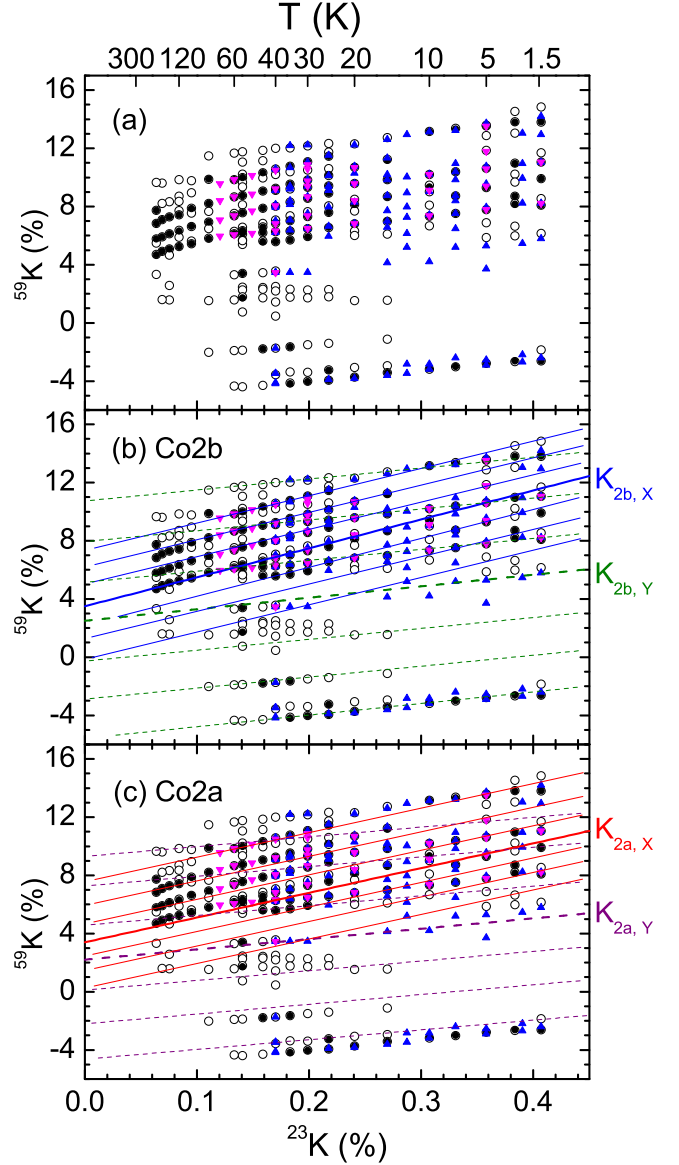}
\caption{(Color online) (a) The global summary of all our magnetic shift
data for the various singularities of the fast relaxing cobalt spectra,
plotted versus the isotropic component of the $^{23}$Na NMR shift. The most
pronounced singularities in the experimental spectra are marked by filled
symbols, whereas weaker singularities are displayed with open symbols.
Different symbols correspond to spectra for differently prepared samples of
the same phase. (b) Here, the same data are reported, but with lines drawn
which underline the central line and 6 quadrupolar satellites of $^{59}$Co
in the Co2b positions for $H_{0}\parallel X$ (blue full lines) and $%
H_{0}\parallel Y$ (green dashed lines) directions; (c) On the same data
again, we have drawn the seven $^{59}$Co NMR lines which correspond to the
Co2a positions for $H_{0}\parallel X$ (red full lines) and $H_{0}\parallel Y$
(magenta dashed lines) directions.}
\label{fig:KCoSingularities}
\end{figure}

We measured then the $^{59}$Co NMR spectra at different temperatures in
swept field mode and at each temperature pointed the positions (field
values) of all singularities in the spectra of fast relaxing Co2. After that
we converted them into field shift values using Eq.~\ref{eq:Kdef} and
plotted them versus $^{23}K(T)$ taken at the measurement temperature. Each
spectrum taken at a given $T$ then allows to report all the data points on a
vertical line in Fig.~\ref{fig:KCoSingularities}a.

As one can see, some singularities for the fast relaxing Co2 appear well
organized around extremal values of the shift, with 3+3 satellites spaced by
the quadrupole splitting $\nu _{\alpha }$ - in Fig.~\ref{fig:KCoSingularities}b we show two such sets of singularities by lines. Initially we assumed that
these two sets of singularities correspond to the two cobalt sites: one with
a large shift in $H_{0}\perp c$ direction and the other a small shift, but
this assumption did not allow us to explain all the data.

Unexpectedly we found that the values of quadrupole splitting $\nu_{\alpha}$
for these two sets of singularities correspond to the $\nu_{X}$ and $%
\nu_{Y} $ values of the Co2b site (the corresponding $\nu_{Q}$ and $\eta$
values were determined by NQR with high presicion\cite{H67NQRprb}).
Therefore the main conclusion of Fig.~\ref{fig:KCoSingularities}b is that we
have a large anisotropy of in-plane magnetic shift for the dominant Co2b
site with distinct values of $K_{XX}$ and $K_{YY}$.

As the number of Co2a sites in the unit cell of the Na$_{2/3}$CoO$_{2}$
phase is twice smaller than that of Co2b sites, the intensity of their NMR
signal is also twice smaller. This makes more difficult to resolve the Co2a
site in the $H_{0}\perp c$ spectra. As one can see in Fig.~\ref%
{fig:KCoSingularities}b many singularities are located in between the
identified Co2b singularities. Using them and the values of $\nu _{Q}$ and $%
\eta$ for Co2a sites,\cite{H67NQRprb} one can suggest two sets of
singularities for the Co2a sites - see Fig.~\ref{fig:KCoSingularities}c. And
in fact the Co2a sites exhibit the same in-plane magnetic shift anisotropy
as the Co2b sites - just with slightly smaller values of the magnetic shifts
and quadrupole splittings.

\begin{figure}[tbp]
\center
\includegraphics[width=1\linewidth]{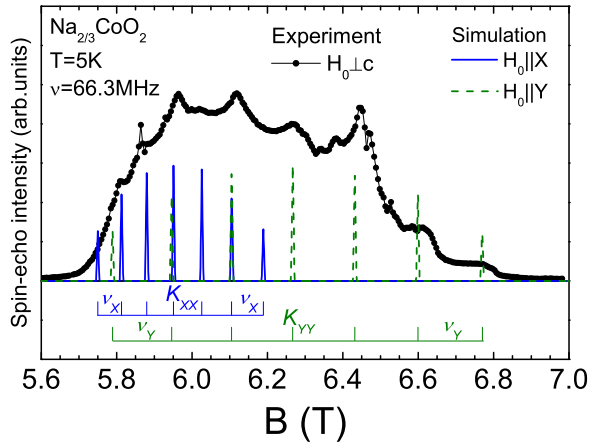}
\caption{(Color online) The $^{59}$Co NMR spectrum of fast relaxing Co2 is
shown by full dots joined by a continuous line. The positions of the seven
lines estimated from the full data analysis are shown for $H_{0}\parallel X$
(blue solid line) and $H_{0}\parallel Y$ (green dashed line). The
correspondence with the singularities in the experimental spectrum is clear.
The position of these two set of singularities is then determined by these
extremal lines while most of the intensity in the spectrum comes from the
particles with intermediate orientations. The large singularities around
6.4~Tesla comes from imperfect compensation of the large signal of the slow
realaxing Co1 sites.}
\label{fig:CoABexplanation}
\end{figure}

Therefore the fast relaxing cobalt $H_{0}\perp c$ spectra can be finally
explained in a simple manner. In Fig.\ref{fig:CoABexplanation} we show the
simulated set of NMR lines of Co2b for the two outermost cases for which $%
H_{0}\parallel X$ and $H_{0}\parallel Y$. These two spectra form the main
singularities in the cobalt spectra while most of its intensity comes from
the particles with intermediate orientations. The large difference in the $%
K_{XX}$ and $K_{YY}$ values makes the cobalt $H_{0}\perp c$ spectra totally
asymmetric.

% Produces the bibliography via BibTeX.
\bibliography{H67CoShift}

\end{document}